\documentclass{article}

\usepackage{arxiv}
\usepackage[utf8]{inputenc}
\usepackage{tabularx}
\usepackage[figuresright]{rotating}
\usepackage{amsmath,amssymb,bbm}
\usepackage{relsize}
\usepackage{graphicx}
\usepackage{arydshln}
\usepackage{booktabs}
\usepackage{multirow}
\usepackage{natbib}
\usepackage{float}
\usepackage{appendix}
\usepackage{algorithm}
\usepackage{makecell}
\usepackage{setspace}

\usepackage[noend]{algpseudocode}
\DeclareMathOperator{\Tr}{Tr}
\newcommand{\te}[1]{\text{#1}}
\newcommand{\ti}[1]{\textit{#1}}

\newcommand{\E}[2]{\mathbbm{E}_{#1}\left[{#2}\right]}

\newcommand\numberthis{\addtocounter{equation}{1}\tag{\theequation}}

\title{Variational Inference of Dynamic Factor Models with Arbitrary Missing Data}

\author{
  Erik Sp{\aa}nberg \\
  Department of Statistics, Stockholm University \\ SE-106 91, Stockholm \\ E-mail: erik.spanberg@stat.su.se
}
\bibpunct[, ]{(}{)}{;}{a}{,}{,}

\begin{document}

\maketitle
\onehalfspacing
\begin{abstract} 
    Dynamic factor models are often estimated by point-estimation methods, disregarding parameter uncertainty. We propose a method accounting for parameter uncertainty by means of posterior approximation, using variational inference.  Our approach allows for any arbitrary pattern of missing data, including different sample sizes and mixed frequencies. It also yields a straight-forward estimation algorithm absent of time-consuming simulation techniques. In empirical examples using both small and large models, we compare our method to full Bayesian estimation from MCMC-simulations. Generally, the approximation captures factor features and parameters well, with vast computational gains. The resulting predictive distributions are approximated to a very high precision, almost indistinguishable from MCMC both in and out of sample, in a tiny fraction of computational time.
\end{abstract} 


\section{Introduction}

Analysis of large data sets is a growing field of study. One way of dealing with big data is imagining some latent, unobserved, process driving observable co-movements. Estimating that process can be a practical and operable way of encapsulating the core information in the data. In particular, if we imagine the dimension of the latent process to be small, large data sets may be summarized by a few, but ideally very relevant, factors. Typically, this is the view taken in the dynamic factor model (DFM), making it a popular data analyzing tool \citep[good surveys are given by][]{BaiWang2016, StockWatson2016}. DFMs have become particularly popular in macroeconomic forecasting, not least due to their ability to deal with unsynchronized data sets with missing data in sample endpoints. This can occur, for instance, when different variables have different publication dates, which is a relevant problem in real-time forecasting of the macroeconomic state  \citep[e.g.][]{GiannoneEtAl2008,BanburaEtAl2011,BanburaRunstler2011,HindrayantoEtAl2016}. Some influencial applications of DFMs are given by \cite{Sargent1989}, \cite{StockWatson1989, StockWatson2002a,StockWatson2002b} and \cite{BaiNg2006}. Additionally, applying DFMs to macroeconomic data is arguably merited by economic theory, which often predicts economic shocks to be pervasive and possibly persistent \citep[see][for an overview of big data methods in macroeconomics]{Fuleky2020}. 
The growing popularity of real time applications enhances the need for fast and simple estimation methods in practice. 

Over the last several decades, various estimation methods have been tested and suggested. DFMs where originally proposed by \cite{Geweke1977} and \cite{SargentSims1977}, estimating in frequency domain \citep[see also][]{GewekeSingleton1981}. Maximum likelihood (ML) estimation in time domain were developed by \cite{EngleWatson1981} utilizing state space representation and the Kalman filter.  \cite{WatsonEngle1983} speed up ML-estimation by means of the Expectation-Maximization (EM) algorithm by \cite{DempsterEtAl1977}. \par 

Of particular interest for this paper, \cite{BanburaetAl2014} construct a reasonably fast EM-algorithm for ML-estimation in case of arbitrary missing data, by extending the results of \cite{RubinThayer1982} and \cite{WatsonEngle1983}. 
Their algorithm allows for a wide range of unbalanced and messy data patterns, including different sample sizes, imputation errors and mixed frequencies. In turn, \cite{Spanberg2021} broaden the approach to encompass maximum a-posteriori estimation. These are practical tools in dealing with real macroeconomic data, attending specific data availability while maintaining a viable ability of updating estimates frequently. They are, however, limited to point estimation.  

In fact, all of the estimation methods above deal with parameter point estimation, disregarding parameter uncertainty. Consequently, they might misrepresent uncertainties in general, including those surrounding predictions. Parameter uncertainty can often be addressed by more computationally demanding simulation techniques such as bootstrapping. Another approach is Bayesian inference, commonly also by computationally demanding simulations.
\cite{OtrokWhiteman1998} implement Bayesian estimation of DFMs using a Markov chain Monte Carlo (MCMC) algorithm to simulate from the posterior distribution. \cite{DelNegroOtrok2008} also considers Bayesian estimation, with an extension of time varying factor loadings and stochastic volatility in idiosyncratic components. These algorithms are much more time-consuming than, for instance, ML estimation of \cite{BanburaetAl2014}, which impedes their practical utility in fast-paced real-time forecasting settings. 

This paper aims to confront the gap between fast point parameter estimation of DFMs and more time-consuming Bayesian estimation. We apply a variational inference (VI) method, the structured mean-field approximation (see Section \ref{sec.2.1}), to find simplified analytical expressions of blockwise-marginal posterior distributions, yielding an estimation algorithm which is both fast and considers parameter uncertainty. To our knowledge, this is a novel approach for standard DFMs. The algorithm bears many similarities to the EM-algorithm of \cite{BanburaetAl2014}. It reaps comparable benefits both in speed and in allowing for arbitrary patterns of missing data, but with the added benefit of estimated parameter densities. Furthermore, we include parameter shrinkage by prior distributions. Combining dimension reduction with parameter shrinkage increases the versatility of the model, including the possibility of keeping parsimony with larger number of factors and increased number of lags in the dynamical specification. 

VI is often applied out of necessity to deal with complex models that are unfit for less approximative methods. DFMs are still in many cases fit for ordinary MCMC methods, however they might not be computationally practical for forecast institution with daily data updates. This problem is increased as DFMs scale up, in terms of variables, lags and factors. Forecasters may consequently settle for point-estimation only to adhere their time constraints. 

By our method they can settle for more. It is much faster than ordinary MCMC and require much less memory to store. Even for larger DFMs, a forecaster could run their daily update in seconds or a few minutes on a standard laptop. 

The remainder of the paper is structured as follows: Section 2 introduces the econometric framework, including model specifications, prior distributions and the VI-algorithm; Section 3 compares the posterior approximation to simulated posteriors by MCMC-algorithm for real data examples; and Section 4 concludes.  
\section{Econometric framework} \label{sec.2}
Let $\ti{y}_t = [\ti{y}_{1,t}, ..., \ti{y}_{n,t}]', n \in \mathbbm{N}$, be a $n$-length column vector from a stochastic process $\{\ti{y}_t, t\in \mathbbm{Z}\}$. Suppose we observe (possibly partially) a realisation of the process for time steps $t=1, ..., T$. The observation equation is given by
\begin{align}
    \ti{y}_t &= \Lambda_0 f_t +  \Lambda_1 f_{t-1} + ... + \Lambda_p f_{t-p} + \epsilon_t, \quad &\epsilon_t \sim \mathcal{N}\left(0, \Sigma_\epsilon \right), \label{yeq}
\end{align}
where $f_t = [f_{1,t}, ..., f_{r,t}]', r \in \mathbbm{N}$, is a $r$-length column vector of dynamic latent factors and $\epsilon_t$ is a $n$-length column vector of \textit{idiosyncratic components}. The factors load onto the time series dynamically, defined by factor loading parameters collected in $p + 1$ separate $(n \times r)$-matrices \{$\Lambda_j: j=0,...,p, \: p \in \mathbbm{N}$\}. Idiosyncratic components are assumed independent with diagonal covariance matrix $\Sigma_\epsilon = \text{diag}\left(\sigma^2_1, ..., \sigma^2_n\right)$. The forced diagonal property of $\Sigma_\epsilon$ defines \eqref{yeq} as an \textit{exact} factor model, in contrast to an \textit{approximate} model with non-diagonal idiosyncratic covariance. For the approximate model, \cite{ForniEtAl2000,ForniEtAl2004,ForniEtAl2005} show consistency of dynamical principal component estimation, as sample size and cross-section size goes to infinity \citep[see also two-step estimator by][]{DozEtAl2011}. 

Further, we assume factors to follow a vector autoregressive process (VAR) according to 
\begin{align}
    f_t &= \Phi_1 f_{t-1} + \Phi_2 f_{t-2} + ... + \Phi_{p+1} f_{t-p-1} + \ti{u}_t, \quad &\ti{u}_t \sim \mathcal{N}\left(0, I_r \right), \label{feq}
\end{align}

where $\te{u}_t$ is a $r$-length column vector of standard normal distributed residuals and $\{\Phi_{j+1}: j=0,...,p, \: p \in \mathbbm{N}\}$ are $(r \times r)$-matrices of parameters. We can write \eqref{yeq}-\eqref{feq} in a state space representation by stacking vectors and matrices:
\begin{align}
    \ti{y}_t &= \Lambda F_t + \epsilon_t, \quad &\epsilon_t \sim \mathcal{N}\left(0, \Sigma_\epsilon \right), \label{yeqSS} \\
    F_t &= \widetilde{\Phi} F_{t-1} + \begin{bmatrix} I_r \\ 0_{rp \times r} \end{bmatrix} \ti{u}_t, \quad &\ti{u}_t \sim \mathcal{N}\left(0, I_r \right), \label{feqSS}
\end{align}
where 
\begin{align*}
\widetilde{\Phi} &= \left[
    \begin{array}{cc}
      \multicolumn{2}{c}{\Phi},  \\ \hdashline[2pt/2pt]
       \multicolumn{1}{c}{\underset{rp \times rp}{I}} & \multicolumn{1}{;{2pt/2pt}c}{\underset{rp \times r}{0}}
    \end{array}\right], \\
F_t &= \left[f_t' \; f_{t-1}' \;  ... \; f_{t-p}' \right]', \\
\Lambda &= \left[\Lambda_0 \; \Lambda_1 \; ... \; \Lambda_p \right], \\
\Phi &= \left[\Phi_1 \; \Phi_2 \; ... \; \Phi_{p+1} \right].
\end{align*} 

Introducing terminology in this context, we call \eqref{yeqSS} the \textit{observation equation}, with \textit{observational vector} $\te{y}_t$, and \eqref{feqSS} the \textit{state equation}, with \textit{state vector} $F_t$. Further, $\Lambda$ is the \textit{factor loading matrix}, $\widetilde{\Phi}$ the \textit{transition matrix} and $\ti{u}_t$ the \textit{state residual}. $\epsilon_t$ is still called idiosyncratic component, with idiosyncratic covariance matrix $\Sigma_\epsilon$.  The \textit{common component} is given by $X_t = \Lambda F_t$, where $F_t$ can be regarded as $s=r(p+1)$ static factors, which are restricted by a particular dynamic structure. 

All parameters in this system are given prior distributions. Let $\lambda_i'$ denote the $i$th row of $\Lambda$, i.e. factor loadings corresponding to the $i$th variable. We define prior distributions for $\Lambda$ and $\Sigma_\epsilon$: 
\begin{align}
    &\lambda_i|\sigma^2_i \sim \mathcal{N}\left(0, \sigma_i^2 V\right), \quad i=1, ..., n, \label{lambdaprior} \\
    &\sigma^2_i \sim \text{Scaled-Inv-}\chi^2\left(\nu_i, \tau^2_i\right), \quad i=1, ..., n, \label{sigmaprior}
\end{align}
where Scaled-Inv-$\chi^2\left(\cdot, \cdot\right)$ denotes the Scaled-Inverse-chi-square distribution \citep[see e.g.][Appendix A]{GelmanEtAl2013}. Shrinkage on individual $\lambda_i$'s is defined by hyperparameters collected in  positive semidefinite $(s \times s)$-matrix $V$. Prior distribution of $\sigma^2_i$ is defined by prior degrees of freedom $\nu_i > 0$ and prior scale parameter $\tau^2_i > 0$. \par  If $\{F_t: t=1,...,T\}$ was observed (which it's not), estimation of $\Lambda$ and $\Sigma_\epsilon$ would reduce to ordinary Bayesian linear regression; \eqref{lambdaprior}-\eqref{sigmaprior} constitute conjugate priors to a linear regression model. Similarly, observed factors would reduce $\eqref{feq}$ to a regression with our $\Phi$-prior as conjugate:
\begin{align*}
    \Phi \sim \mathcal{MN}_{r,s}\left(0, I_r, W\right),
\end{align*}
where $\mathcal{MN}_{.,.}(\cdot,\cdot,\cdot)$ denotes the Matrix Gaussian distribution \citep[see e.g.][Chapter 2]{GuptaNagar1999}. The parameter shrinkage is defined by hyperparameters collected in postive semidefinite $(s \times s)$-matrix $W$. Lastly, we initialize the state space system by introducing a prior on origin state vector 
\begin{align*}
    F_0 \sim \mathcal{N}\left(0, \Sigma_{F_0}\right),
\end{align*}
where initial state covariance $\Sigma_{F_0}$ is a positive definite $(s \times s)$-matrix.  

\subsection{Variational inference} \label{sec.2.1}

VI approximates posterior distributions arising in Bayesian inference. An introduction to VI is given by \cite{JordanEtAl1999}, and a review by \cite{BleiEtAl2017}. Consider some set of observed data $\mathcal{D}$ and set of unknowns $\mho$ (e.g. static parameters or time-varying latent variables). Bayes' Theorem states 
\begin{align*}
    p(\mho|\mathcal{D}) = \frac{p(\mathcal{D}|\mho)p(\mho)}{p(\mathcal{D})} \propto  p(\mathcal{D}|\mho)p(\mho),
\end{align*}
where $p(\mathcal{D}|\mho)$ is the likelihood function, $p(\mho)$ the prior density and $p(\mathcal{D})$ the marginal likelihood. In Bayesian inference, the posterior distribution $p(\mho|\mathcal{D})$ is generally of particular interest, either as a target itself or as an aide in some decision problem. However, in most realistic problem formulations, it's not given to us in a neat and directly accessible form. In fact, it's often intractable. The most common approach is to explore the posterior distribution by MCMC-simulation, which can be computationally heavy and time-consuming, especially when $\mho$ is highly dimensional. 

VI provides a way to reduce computational costs by approximating posterior distributions to more accessible forms. We define \textit{variational density} $q(\mho)$ as a probability density function in some set of densities $Q$.  VI formulates an optimization problem: 
\begin{align}
    q^\star(\mho) = \underset{q \in Q}{\text{arg min}} \: \text{KL}\left(q(\mho)||p(\mho|\mathcal{D})\right), \label{VIopt}
\end{align} 
where $\text{KL}\left(\cdot||\cdot\right)$ is the Kullback-Leibler (KL) divergence \citep[see][]{KullbackLeibler1951}. In other words, $q^\star(\mho)$ is the member of density set $Q$ which minimizes the KL divergence to the posterior density. Minimizing KL is the same as maximizing the so called \textit{evidence lower bound} (ELBO) given by
\begin{align}
    \text{ELBO} = \int q(\mho) \ln \frac{p(\mathcal{D}|\mho)p(\mho)}{q(\mho)}d\mho = \E{q}{\ln p(\mathcal{D}, \mho)} - \E{q}{\ln q(\mho)}.\label{ELBO}
\end{align} 

If we can formulate $Q$ with computationally accessible members that entails a viable approximation of $p(\mho|\mathcal{D})$, and if optimization \eqref{VIopt} is tractable and not unreasonably difficult, $q^\star(\mho)$ satisfy our goals.

There are several ways to choose $Q$. The main approaches are \textit{fixed-form} and \textit{mean-field}. By fixed-form approximation, $Q$ is chosen to be a particular class of distributions, e.g. multivariate Gaussian. In mean-field approximation on the other hand, distributions are not restricted to a particular class. Instead, we construct $Q$ as all possible densities satisfying some independency restrictions. Specifically, we define a partition $\mho = \left\{\mho_1, \mho_2, ..., \mho_m \right\}$ with a restricting factorizing equality $q(\mho) = q(\mho_1)q(\mho_2)...q(\mho_m)$. When the partition is defined by blocks of unknowns, we speak of a \textit{structured mean-field} (SMF) approximation. 

The optimal variational densities by SMF are given by 
\begin{align*}
    q^\star\left(\mho_1\right) &\propto \exp\left\{\E{q(\mho \backslash \mho_1)}{\ln p(\mathcal{D}, \mho)}\right\}, \\
    q^\star\left(\mho_2\right) &\propto \exp\left\{\E{q(\mho \backslash \mho_2)}{\ln p(\mathcal{D}, \mho)}\right\}, \\
                               &\: \: \vdots \\
    q^\star\left(\mho_m\right) &\propto \exp\left\{\E{q(\mho \backslash \mho_m)}{\ln p(\mathcal{D}, \mho)}\right\}.  \numberthis \label{SMFsteps}                  
\end{align*}
In other words: log optimal variational densities for individual unknown blocks are given by taking the expectation of log joint density of $\mathcal{D}$ and $\mho$, with respect to all other unknown blocks.

There is no obvious fixed-form class choice for DFMs, not least because our unknowns include both time-varying states and static parameters. Rather, we propose SMF approximation by \eqref{SMFsteps}, with an independency restriction between common factors and static parameters. 

\subsection{Variational inference of DFMs} 

Our target unknowns are $\mho = \{\theta, F\}$ including parameters $\theta = \{\Lambda, \Sigma_\epsilon, \Phi\}$ and state vectors $F = \{F_t: t=0,...,T\}$. We want to approximate the posterior distribution by a member of a family of distributions $Q$, satisfying the equality $q(\theta, F) = q(\theta)q(F)$. Missing data is allowed and we define data $\mathcal{D} = Y_A$ as the set of all available observations in $\{\ti{y}_t: t=1,...,T\}$. The optimal variational densities by SMF is thus given by
\begin{align}
    q\left(\theta\right) &\propto \exp\left\{\E{q(F)}{\ln p\left(\theta, F, Y_A\right)}\right\}, \\
    q\left(F\right) &\propto \exp\left\{\E{q(\theta)}{\ln p\left(\theta, F, Y_A\right)}\right\},
\end{align}

where we make use of the log joint density
\begin{align*} 
    \ln p\left(Y_A, F, \theta\right) = \ln p\left(Y_A| \theta, F \right) + \ln p\left(F| \theta \right) + \ln p\left(\theta \right),
\end{align*}
 in which $p\left(Y_A| \theta, F \right)$ is the likelihood function, $p(F|\theta)$ is the parameter conditional factor density and $p(\theta)$ is the prior density of $\theta$.

Due to the diagonal property of $\Sigma_\epsilon$, we can factorize and directly integrate out missing observations to construct the likelihood function (see Appendix \ref{sec.A1}). Let's denote indicator variables by
\begin{align*}
    a_{i,t} = \begin{cases} 1 \quad \text{ if } \ti{y}_{i,t} \text{ is available} \\
    0 \quad \text{ if } \ti{y}_{i,t} \text{ is missing} \end{cases} 
\end{align*}
and the number of available observations for time series variable $i$ as $T_i = \sum_{t=1}^T a_{i,t}$. The likelihood function is given by 
\begin{align*}
    p\left(Y_A|\theta,F\right) = \prod_{i=1}^n\left(\left(\frac{1}{2\pi\sigma_i^2}\right)^{T_i} \exp\left\{-\sum_{t=1}^T \frac{a_{i,t}}{2\sigma_i^2}\left(\ti{y}_{i,t} - \lambda_i' F_t\right)^2 \right\} \right) \numberthis \label{likfun}
\end{align*}
and the parameter conditional factor density is given by
\begin{align*}
    p(F|\theta) &= (2\pi)^{s/2}\det\left(\Sigma_{F_0}\right)^{-1/2}\exp\left\{-\frac{1}{2}F_0'\Sigma_{F_0}^{-1}F_0\right\} \\ &\quad \times \prod_{t=1}^T\left((2\pi)^{s/2}\exp\left\{-\frac{1}{2}\left(F_t - \widetilde{\Phi}F_{t-1}\right)'\left(F_t - \widetilde{\Phi}F_{t-1}\right)\right\}\right). \numberthis \label{condFdens}
\end{align*} 
Taking logs of \eqref{likfun}-\eqref{condFdens} and adding log-prior densities gives
\begin{align*}
    \ln p\left(Y_A, F, \theta \right) &= C - \frac{1}{2}\sum_{i=1}^n T_i \ln \sigma_i^2 - \sum_{i=1}^n\sum_{t=1}^T \frac{a_{i,t}}{2\sigma_i^2}\left(\ti{y}_{i,t} - \lambda_i' F_t\right)^2 \\
    &- \frac{1}{2}F_0' \Sigma_{F_0}^{-1} F_0 - \frac{1}{2}\sum_{t=1}^T\left(F_t - \widetilde{\Phi}F_{t-1}\right)'\left(F_t - \widetilde{\Phi}F_{t-1}\right) \\
    &-\frac{s}{2}\sum_{i=1}^n \ln \sigma_i^2  -\frac{1}{2} \sum_{i=1}^n \frac{1}{\sigma^2_i} \lambda_i'V^{-1}\lambda_i \\ 
    &- \sum_{i=1}^n \left(1 + \nu_i/2\right) \ln \sigma^2_i - \sum_{i=1}^n \frac{\nu_i \tau^2_i}{2\sigma^2_i}  -\frac{1}{2}\Tr\left(\Phi W^{-1} \Phi'\right), \numberthis \label{jointdens}
\end{align*}
where $C$ is a constant in terms of $\theta$ and $F$.
\newpage 
\subsubsection{Variational density of static parameters}
In this section we describe the variational density of $\theta$. As $\{\Lambda, \Sigma_\epsilon\}$ shares no common terms with $\Phi$ in \eqref{jointdens}, we can deduce that
\begin{align*}
    q\left(\theta\right) = q\left(\Lambda, \Sigma_\epsilon\right)q\left(\Phi\right).
\end{align*}
This has an intuitive explanation. The posterior dependency between parameter blocks $\{\Lambda, \Sigma_\epsilon\}$ and $\Phi$ is contingent upon their respective individual dependencies of $F$. If we force independence between $\theta$ and $F$, independence between parameter blocks is subsequent. 

The variational density for $\{\Lambda, \Sigma_\epsilon\}$ is derived in Appendix \ref{sec.A2} and shown to be equation-wise independent Gaussian-Scaled-Inverse-Chi-Square\footnote{Sometimes referred to Gaussian-Inverse-Gamma, but with a different parametrisation.} according to \begin{align}
    q\left(\Lambda, \Sigma_\epsilon\right) = \prod_{i=1}^n q\left(\lambda_i|\sigma^2_i\right)q\left(\sigma^2_i\right) = \prod_{i=1}^n \mathcal{N}\left(\lambda_i\Big|\mu_{\lambda_i}, \sigma^2_i \Sigma_{\lambda_i} \right)  \text{Scaled Inv-}\chi^2\left(\sigma^2_i\Big|\nu_{\sigma_i}, \tau^2_{\sigma_i} \right), \label{lambdasigmadens}
\end{align} where \begin{align}
       \mu_{\lambda_i} &= \left[\sum_{t=1}^T a_{i,t}\E{q(F)}{F_t F_t'} + V^{-1}\right]^{-1}\left[\sum_{t=1}^T \E{q(F)}{F_t}a_{i,t}\ti{y}_{i,t}\right], \label{mulambda} \\
       \Sigma_{\lambda_i} &= \left[\sum_{t=1}^T a_{i,t}\E{q(F)}{F_t F_t'} + V^{-1}\right]^{-1}, \label{Sigmalambda} \\
       \nu_{\sigma_i} &= \nu_i + T_i,  \label{nusigma}\\
       \tau^2_{\sigma_i} &= \frac{1}{\nu_{\sigma_i}}\left(\nu_i\tau^2_i + \sum_{t=1}^T a_{i,t}\ti{y}_{i,t}^2 - \mu_{\lambda_i}'\Sigma_{\lambda_i}^{-1}\mu_{\lambda_i}\right). \label{tausigma}
\end{align}

Expressions \eqref{mulambda}-\eqref{tausigma} are very similar to posterior parameters in ordinary Bayesian linear regression with conjugate priors. The difference lies in the exchange of regressors with expectations, specifically factor moments $\E{q(F)}{F_t}$ and $\E{q(F)}{F_tF_t'}$. Prior covariance $V$ can be interpreted as a shrinkage parameter matrix; the smaller $V$ is, the stronger shrinkage towards $\mu_{\lambda_i} = 0$. We can also interpret prior degrees of freedom $\nu_i$ as a prior number of observations of $\sigma^2_i$, taking the value of prior scale $\tau^2_i$. The larger $\nu_i$ is, the stronger prior information that $\sigma^2_i = \tau^2_i$. Perhaps somewhat abstractly, $\nu_i$ does not have to be a whole number. 

 Worth noting is that if $V^{-1}=0$, \eqref{mulambda} is analogous to ML estimation M-step by \cite{BanburaetAl2014} (equation (11), p.138 therein).\footnote{Correspondingly, if $V^{-1} \neq 0$, the expression is analogous to maximum a-posteriori estimation of \cite{Spanberg2021}.} Similarily, if $\nu_i = 0$, \eqref{tausigma} is analogous to ML estimation M-step of idiosyncratic variances.\footnote{The variance estimator expression differs additionally from M-step by \cite{BanburaetAl2014} (equation (12), p.138 therein). Their estimator includes taking the expectation over missing observations, which yields an unnecessarily complicated and computationally slower expression, as missing observations can be integrated out preemptively. This was pointed out by \cite{Spanberg2021}.}  Although the expressions become analogous, the factor moments within will likely differ. Also, observe that in these cases the priors will be improper and ELBO undefined. 

Furthermore, only terms corresponding to available data are added and multiplied; if there are no available data the estimates reduce to prior parameters. In theory, we could include variables without available observations, although their corresponding parameters would only stay at their prior distributions and have no bearing on factor estimates.

To accommodate notations in coming sections we introduce expected moments in matrix form:
\begin{align*}
     M_\Lambda &\equiv \E{q\left(\Lambda|\Sigma_\epsilon\right)}{\Lambda} =  \begin{bmatrix} \mu_{\lambda_1}' \\ \vdots \\ \mu_{\lambda_n}' \end{bmatrix}, \\
    \Psi^{-1} &\equiv \E{q\left(\Sigma_\epsilon\right)}{\Sigma_\epsilon^{-1}} = \text{diag}\left(\tau^2_{\sigma_1}, ..., \tau^2_{\sigma_n}\right)^{-1}.
\end{align*} 

We derive the variational density of $\Phi$ in Appendix \ref{sec.A3}, which is found in a comparable way. It's given by
\begin{align}
    q(\Phi) = \mathcal{MN}_{r,s}\left(M_\Phi, I_r, \Sigma_\Phi\right), \label{phidens}
\end{align}
where 
\begin{align}
    M_\Phi &=  \left(\sum_{t=1}^T \E{q\left(F\right)}{f_tF_{t-1}'}\right)\left(\sum_{t=1}^T \E{q\left(F\right)}{F_{t-1} F_{t-1}'} + W^{-1}\right)^{-1}, \label{M_Phi}\\ 
    \Sigma_\Phi &= \left(\sum_{t=1}^T \E{q\left(F\right)}{F_{t-1} F_{t-1}'} + W^{-1}\right)^{-1}. \label{Sigma_Phi}
\end{align}

Similar to $q(\Lambda|\Sigma_\epsilon)$,   \eqref{M_Phi}-\eqref{Sigma_Phi} are mirroring a multiple Bayesian linear regression with conjugate priors and known variance, but where the sufficient statistics are exchanged by expectations $\E{q(f)}{F_{t-1}F_{t-1}'}$ and $\E{q(F)}{f_t F_t'}$. Like previously, $W$ can be interpreted as a parameter shrinkage matrix, with smaller values leading to stronger shrinkage towards $M_\Phi=0$. If $W^{-1} = 0$, then \eqref{M_Phi} is analogous to M-step by \cite{BanburaetAl2014} (equation (6), p.137 therein). 

\subsubsection{Variational density of latent states}
In this section we describe the variational density of $F$. It is derived in Appendix \ref{sec.A4}, and given by
\begin{align*}
    q(F) &= C_F\exp\left\{-\frac{1}{2}\sum_{t=1}^T\left(\begin{bmatrix} \ti{y}_t \\ 0_{s \times 1} \end{bmatrix} - \begin{bmatrix} M_\Lambda\\ I_{s} \end{bmatrix}  F_t\right)'\begin{bmatrix}A_t\Psi^{-1} & 0_{n \times s} \\ 0_{s \times n} & \Sigma_t^\theta \end{bmatrix}\left(\begin{bmatrix} \ti{y}_t \\ 0_{s \times 1} \end{bmatrix} - \begin{bmatrix} M_\Lambda \\ I_s \end{bmatrix}  F_t\right) \right.\\
    &\quad \quad  \quad \quad \quad \; \left. - \frac{1}{2}\sum_{t=1}^T \left(F_t - \widetilde{M}_\Phi F_{t-1}\right)'\left(F_t - \widetilde{M}_\Phi F_{t-1}\right) -  \frac{1}{2}F_0'\left(\Sigma_{F_0}^{-1} + r\Sigma_\Phi\right)F_0\right\}, \numberthis \label{Fdens}
\end{align*}
where $C_{F}$ is constant with respect to $F$. \eqref{Fdens} describes a state space system according to:
\begin{align}
    \widetilde{\ti{y}}_t &= \widetilde{M}_\Lambda F_t + \widetilde{\epsilon}_t, \quad \widetilde{\epsilon}_t \sim \mathcal{N}\left(0, \widetilde{\Sigma}_t \right), \label{Fss1} \\
    F_t &= \widetilde{M}_\Phi F_{t-1} + \begin{bmatrix} I_r \\ 0_{rp \times r} \end{bmatrix}\ti{u}_t, \quad \ti{u}_t \sim \mathcal{N}\left(0, I_r\right), \label{Fss2} \\
    F_0 &\sim \mathcal{N}\left(0, \left(\Sigma_{F_0}^{-1} + r\Sigma_\Phi\right)^{-1}\right), \label{Fss3}
\end{align}

where 

\begin{align*}
    \widetilde{\ti{y}}_t &= \begin{bmatrix} \ti{y}_t \\ 0_{s \times 1} \end{bmatrix}, \quad \widetilde{M}_\Lambda = \begin{bmatrix} M_\Lambda \\ I_s \end{bmatrix}, \quad \widetilde{M}_\Phi = \left[
    \begin{array}{cc}
      \multicolumn{2}{c}{M_\Phi}  \\ \hdashline[2pt/2pt]
       \multicolumn{1}{c}{\underset{rp \times rp}{I}} & \multicolumn{1}{;{2pt/2pt}c}{\underset{rp \times r}{0}}
    \end{array}\right], \quad  \widetilde{\Sigma}_t =  \begin{bmatrix} \Psi & 0_{n \times s} \\ 0_{s \times n} & \left(\Sigma_t^\theta\right)^{-1}  \end{bmatrix}, \\
        A_t &= \begin{bmatrix} a_{1,t} & 0 & \dots & 0 \\ 0 & a_{2,t} & \dots & 0 \\ \vdots &  \vdots & \ddots & \vdots \\ 0 & 0 & \dots & a_{n,t} \end{bmatrix} \text{ and } \Sigma^\theta_t = \begin{cases}   \sum_{i=1}^n a_{i,t} \Sigma_{\lambda_i} + r\Sigma_\Phi\quad &\text{ if } t=1, ..., T-1 \\ \\
     \sum_{i=1}^n a_{i,T} \Sigma_{\lambda_i} \quad &\text{ if } t=T. \end{cases}
\end{align*}

More in detail, \eqref{Fss1}-\eqref{Fss3} can be described as a DFM with factor loadings $M_\Lambda$, idiosyncratic covariance $\Psi$ and transition matrix $\widetilde{M}_\Phi$, but where the observation vector is augmented by a $s$-length 0-vector. We take into account variational covariances of $\Lambda$ and $\Phi$ by including $(\Sigma^\theta_t)^{-1}$ in the observational covariance matrix, specified onto the added $0$-observations. 

$\Sigma^\theta_t$ is a sum of variational covariances for $\Phi$ and $\lambda_i$'s corresponding to available data, and in a sense a measurement of parameter uncertainty at the particular time step. The larger this sum becomes, the smaller covariance around a $0$-vector. 

There is some intuition here. If $q(\Lambda, \Phi)$ is tight, then $q(F)$ will be closely described by a DFM with fixed static parameters at first moments $M_\Lambda$, $\tilde{M}_\Phi$ and $\Psi$. But if $q(\Lambda, \Phi)$ is wide, the first moment $\E{q(F)}{F}$ will be pushed, with higher certainty, towards the unconditional mean of the process: $0$. The uncertainty of $\theta$ acts as an extra shrinkage on $F$. Compare this to E-step by \cite{BanburaetAl2014} (equation (10), p.137 therein), in which the state space system is not augmented. By their method, any parameter uncertainty is discarded in factor estimation, which is a core difference to our procedure.

Additionally, \eqref{Fss1}-\eqref{Fss3} can be simplified into a smaller system by collapsing the observational vector following \cite{JungbackerKoopman2015} \citep[see also][Chapter 6.5]{DurbinKoopman2012}. We collapse the observational vector from $(n + s)$-length to $s$-length, leading to potentially large computational gains. These gains are particularly large when $n>>s$, which is typically the case in DFMs. Also, we counteract some of the increased computations due to augmentation. Appendix \ref{sec.A4.1} derives the collapsed version of \eqref{Fss1}-\eqref{Fss3}. 

Conclusively, the variational density $q(F)$ is the posterior state density, more speedily obtained by running the Kalman filter and smoother over collapsed state space model:

\begin{align}
    \ti{y}^\star_t &=  F_t + \epsilon^\star_t, \quad \epsilon^\star_t \sim \mathcal{N}\left(0, H_t^\star \right), \label{FssC1} \\
    F_t &= \widetilde{M}_\Phi F_{t-1} + \begin{bmatrix} I_r \\ 0_{rp \times r} \end{bmatrix}\ti{u}_t, \quad \ti{u}_t \sim \mathcal{N}\left(0, I_r\right), \label{FssC2} \\
    F_0 &\sim \mathcal{N}\left(0, \left(\Sigma_{F_0}^{-1} + r\Sigma_\Phi\right)^{-1}\right), \label{FssC3}
\end{align}
where 
\begin{align}
   \ti{y}_t^\star &= \left({M_{\Lambda}}'A_t\Psi^{-1} M_\Lambda + \Sigma^\theta_t \right)^{-1}{M_{\Lambda}}'\Psi^{-1}A_t\ti{y}_t,  \label{ystar} \\
   H^\star_t &= \left({M_{\Lambda}}'A_t\Psi^{-1} M_\Lambda + \Sigma^\theta_t \right)^{-1}. \label{Hstar}
\end{align}

From there, we can obtain objects $\E{q(F)}{F_t}$, $\E{q(F)}{f_t F_{t-1}'}$ and $\E{q(F)}{F_t F_t'}$ used in \eqref{mulambda}-\eqref{tausigma} to compute the variational density of $\theta$. 

\subsubsection{SMF iterations}

In previous sections, we have shown the particular functional forms of SMF optimal variational densities. However, we have yet to show how to evaluate the parameters defining these densities, as the variational moments of $\theta$ and $F$ are functions of each other. The most common approach is cooardinate ascent, iteratively updating densities according to
\begin{align*}
    q_j\left(F\right) &\propto \exp\left\{\E{q_{j-1}(\theta)}{\ln p\left(\theta, F, Y_A\right)}\right\}, \\
    q_j\left(\theta\right) &\propto \exp\left\{\E{q_j(F)}{\ln p\left(\theta, F, Y_A\right)}\right\}. \\
\end{align*} As ELBO is convex with respect to $\theta$ and $F$, this is guaranteed to converge to a local optimum \citep[see][]{BoydVandenberghe2004}. Algorithm \ref{algoSMFDFM} describes how to implement this practically in our framework. A natural choice of convergence criteria is relative changes to ELBO between iterations:

\begin{align}
    \text{criteria} = \frac{\text{ELBO}_j - \text{ELBO}_{j-1}}{\frac{1}{2}\left(\left|\text{ELBO}_j\right| + \left|\text{ELBO}_{j-1}\right|\right)}, \label{criteria}
\end{align}

where $\text{ELBO}_j$ is the ELBO for iteration $j$, given and derived in Appendix \ref{sec.A5}. 

Another criteria could be absolute changes in variational moments, but this has some potential drawbacks. First, there are many of them, often in very different scales. Second, moments can change partly due to observationally equivalent parameter and factor rotations, not influencing the ELBO, running the algorithm in unnecessarily many iterations. 

\begin{algorithm} 
  \caption{SMF for DFM} \label{algoSMFDFM}
  \begin{itemize}
    \item[\textbf{1.}] \textbf{Initialization} 
    \begin{itemize}
     \item[a)] Specify numbers of factors $r$ and lag-length $p$. 
     \item[b)] Specify hyperparameters $\Sigma_{F_0}$, $V$, $W$, $\nu_i$ and $\tau_i$ $\forall i = 1, ..., n.$
     \item[c)] Set tolerance level. 
     \item[d)] Set starting parameters $M_\Phi^{(0)}, \Sigma_\Phi^{(0)}, \Psi^{(0)}, M_\Lambda^{(0)}$ and  $\Sigma_{\lambda_i}^{(0)}, \forall i=1,...,n$.
     \item[e)] Set starting index $j \leftarrow 0$.
    \end{itemize}
    \item[\textbf{2.}] \textbf{Updating densities}
    
    \textbf{while} criteria $>$ tolerance level
        \begin{itemize} 
        \item[a)] $j \leftarrow j + 1$
        \item[b)] Update $q_j(F)$ by running Kalman filter and smoother over state space model \eqref{FssC1}-\eqref{FssC3} with parameters $M_\Phi^{(j-1)}$, $\Sigma_\Phi^{(j-1)}$, $\Psi^{(j-1)}$, $M_\Lambda^{(j-1)}$ and  $\Sigma_{\lambda_i}^{(j-1)}, \forall i=1,...,n$.
        \item[c)] Update $q_j(\theta)$ by updating parameters $M_\Phi^{(j)}$, $\Sigma_\Phi^{(j)}$, $\Psi^{(j)}$, $M_\Lambda^{(j)}$ and  $\Sigma_{\lambda_i}^{(j)}$  $\forall i=1,...,n$, using $\E{q_j(F)}{F_t}, \E{q_j(F)}{f_t F_{t-1}'}, \E{q_j(F)}{F_t F_t'}$ and $\E{q_j(F)}{F_0 F_0'}, \forall t=1, ..., T$, given from Kalman smoother in previous step. 
        \item[d)] Update criteria.
        \end{itemize}
        \textbf{end while} 
  \end{itemize}
\end{algorithm}

\subsection{Prior parameters} \label{sec.2.3}
There are several ways to define the prior densities. One relevent suggestion is Minnesota style prior commonly used in Bayesian Vector Autoregressions  \citep[see e.g.][]{Litterman1986, SimsZha1998, BanburaetAl2010, Karlsson2013}. DFM-versions have also been used, for instance by \cite{Spanberg2021}. The suggestion is
\begin{align*}
    V^{-1} &= \eta_\Lambda J_\Lambda, &\quad  J_\Lambda &= \begin{bmatrix} 1 & 2^{\ell_\Lambda} & ... & (p+1)^{\ell_\Lambda} \end{bmatrix} \otimes I_r, \quad &\eta_\Lambda, \eta_\Phi > 0, \\
    W^{-1} &= \eta_\Phi J_\Phi, &\quad  J_\Phi &=  \begin{bmatrix} 1 & 2^{\ell_\Phi} & ... & (p+1)^{\ell_\Phi} \end{bmatrix} \otimes I_r,  \quad &\ell_\Lambda, \ell_\Phi > 1, \numberthis \label{lambdaphiprior}
\end{align*}
where $\eta_\Lambda$ and $\eta_\Phi$ is the overall shrinkage and $\ell_\Lambda$ and $\ell_\Phi$ is the lag-decay for $\Lambda$ and $\Phi$, respectively. The larger overall shrinkage, the greater shrinkage of upon all parameters in respective block, whereas the larger lag-decay, the greater the relative shrinkage upon parameters corresponding to higher lags of $f_t$. The approach allows for adding factor lags with reduced risk of overfitting. It can be interpreted as a prior mean of $\{\ti{y}_t, f_t\}_{t=1}^T$ begin independent gaussian white noise processes, where we assign higher prior density that factors influence observations closer in time. We also have $\nu_i$ prior observations that the idiosyncratic component $\epsilon_{i,t}$ have variance $\tau^2_i$. If $\ti{y}_{i,t}$ is standardized to standard deviation 1, than $\tau^2_i=1$ is consistent with the prior mean of $\Lambda$. 

The hyperparameters $\eta_\Lambda, \eta_\Phi, \ell_\Lambda, \ell_\Phi, \nu_i$ and $\tau_i^2$, have to be assigned by the researcher. Examples of data driven approaches could be running Algorithm \ref{algoSMFDFM} several times, finding hyperparameters minimizing some predictive loss or maximizing ELBO over a training sample. 

\subsection{Factor rotations} \label{sec.2.4}
In general, factors are not uniquely identified by data. \cite{BaiWang2014} prove the identifying conditions for a DFM with specifications \eqref{Fss1}-\eqref{Fss3}, and that it is observationally equivalent under a specific set of rotations \citep[see also][for a specific Bayesian treatment]{BaiWang2015}. We define a particular rotation matrix 
\begin{align*}
    \widetilde{R} = \begin{pmatrix}R & 0 & \dots & 0 \\ 0 & R & \dots & 0 \\ \vdots & \vdots & \ddots & \vdots \\ 0 & 0 & \dots & R   \end{pmatrix}, \quad \text{ where } \widetilde{R} \text{ is a } (s \times s)\text{-matrix and } R \text{ a } (r \times r)\text{-matrix satisfying } RR' = I_r,
\end{align*}

 $F_t$, $\Lambda$ and $\Phi$ can be exchanged by $\widetilde{R}' F_t$, $\Lambda \widetilde{R}$ and $R'\Phi\widetilde{R}$, without affecting the likelihood.  This does not generally mean that the prior, and therefore the posterior, is unaffected by rotation. It will be, however, if we choose $\Sigma_{F_0}$, $V$ and $W$ as diagonal, like for example Minnesota style from Section \ref{sec.2.3}. In that case, the prior will only be rotated in independent, zero-centered and equal variance Gaussians, which are invariant to rotations due to their spherical symmetric property \citep[a characterizing trait of the Gaussian, see][]{Kac1939}. To be clear, not all priors need to have the same variance, but they will within each individual rotational block. As rotations are not made over time dimension, lag-decay can be applied while maintaining rotational invariance. 

It is important to note that Algorithm \ref{algoSMFDFM} only converges to a local optimum. The variational density will be centered at some particular rotation, which can differ depending on starting values. If we are mainly interested in making predictions, this is not particularly relevant in the invariant scheme. And even if we have some rotational preference, we can always make a post-algorithm rotation. 

On the other hand. if $V$ or $W$ are chosen as non-diagonal, some rotations will almost certainly have higher posterior density. This may lead to local convergences to globally sub-optimal rotations. In that case, we might consider exploring the optimization space by running the algorithm with different starting values. One approach could be rotating from a local optimum and start the algorithm again at that spot. 

It is worth pointing out that ordinary MCMC-methods also have a risk of getting stuck at sub-optimal rotations.

\section{Empirical example} 

We employ the methods described in Section \ref{sec.2} and evaluate their approximation precision in empirical examples using Swedish macro-economic data. The data set is described in Appendix \ref{sec.B}, and constitutes of 250 macro-economic variables with different sample sizes and availability, including variables of production, economic surveys, interest rates, employment,  working hours, labor costs, turnovers, consumer and producer prices, real estate statistics, equity prices, airport statistics, tourism, consumption, trade, bankruptcies and exchange rates. We consider both a large DFM using all 250 variables, with $r=2$ and $p=2$, and a small DFM using s subset of 25 variables, with $r=1$ and $p=0$. 

Our comparative benchmark is a MCMC-method, where we use Gibbs Sampling \citep[see example][]{GelmanEtAl2013} to find the posterior $p\left(F,\theta|Y_A\right)$, given by iteratively drawing from full conditional distributions

\begin{align*}
    F^{(d)} &\sim p\left(F\Big|Y_A, \theta^{(d-1)}\right), \\
    \theta^{(d)} &\sim p\left(\theta\Big|Y_A, F^{(d)}\right), \\
    d &= 1, ..., D,
\end{align*}

where we have number of draws D = 200 000, with first 10 \% draws as "burn-in" draws (i.e. not included in the comparison). $p\left(\theta|Y_A, F^{(d)}\right)$ is given by ordinary Bayesian linear regression, whereas we can draw from $p\left(F|Y_A, \theta^{(d-1)}\right)$ using Simulations smoother of \cite{FruhwirthSchnatter1994}  \citep[see also][]{CarterKohn1994}. \footnote{We also considered the method of \cite{DurbinKoopman2002}, but did not find it faster. Also, it had some numerical difficulty dealing with outlier non-stationary explosive draws of $\Phi$.}

To be able to make successful comparisons, we want SMF-approximation and MCMC-benchmark to be found at the same factor rotation, i.e being centered around the same mode. We ensure factor identification by choosing a couple of identifying variables, upon which all factor loadings are set to zero except for one contemporaneous factor.\footnote{Consequentely, all prior and posterior parts corresponding to zero-restricted parameters are removed, for instance in computation of ELBO.} The identifying variables are "Purchasing Managers' Index, Total Manufacturing" for factor 1 (small and large DFM) and "Export, Goods" for factor 2 (large DFM). These variables have a high degree of availability and were greatly connected to factor 1 and 2, respectively, in a simple principal component analysis. The scheme is sufficient to identify factors, all up to, but excluding, sign changes \citep[see e.g.][]{BaiWang2015}. To ensure that the Gibbs Sampler does not jump between modes corresponding to these sign changes, we restrict factor loadings onto identifying variables to be positive, disregarding negative draws. Factor identification is not generally necessary in a practical forecasting setting, but they are necessary to evaluate our approximation precision, as SMF only converges to a local mode (see Section \ref{sec.2.4}). 

All model specifications are the same for MCMC and SMF. We choose prior parameters $\nu_i, \tau^2_i, \eta_\Lambda, \eta_\Phi$ and $\ell_\Phi$ equal to 1; lag-decay $\ell_\Lambda$ and $\ell_\Phi$ equal to 2; and $\Sigma_{F_0} = I_s$. Starting parameters are chosen by a Bayesian linear regression where the regressors are estimated factors from a principal component analysis, in which all missing data have been filled by standard normal random numbers. 

The comparison is made for marginal densities of individual parameters and factors at individual time steps. We also consider in- and out-of sample predictions, which are given by
\begin{align*}
    \hat{\ti{y}}^{(d)}_{i,t} &= \lambda_i^{(d)}F_t^{(d)} + \epsilon^{(d)}_{i,t}, \quad  \epsilon^{(d)}_{i,t} \sim \mathcal{N}\left(0, {\sigma^2_i}^{(d)}\right), \\
    \hat{\ti{y}}^{(d)}_{T+h} &= \lambda_i^{(d)}F_{T+h}^{(d)} + \epsilon^{(d)}_{i,T+h}, \quad \epsilon^{(d)}_{i,T+h} \sim \mathcal{N}\left(0, {\sigma^2_i}^{(d)}\right),
\end{align*}
respectively, where 
\begin{align*}
    F_{T+h}^{(d)} = \Phi^{(d)}F^{(d)}_{T+h-1} + \begin{bmatrix} I_r \\ 0_{rp \times r} \end{bmatrix} \ti{u}_t, \quad &\ti{u}_t \sim \mathcal{N}\left(0, I_r \right).
\end{align*}

In the SMF-case, draws of $\theta$ and $F$ are made from variational densities. 

\subsection{Small DFM}
\begin{figure}[H]
    \centering
      \caption[]{\tabular[t]{@{}l@{}} Small DFM factor, MCMC-posterior and SMF-approximation \\ \\\endtabular}
    \includegraphics[width=1\textwidth]{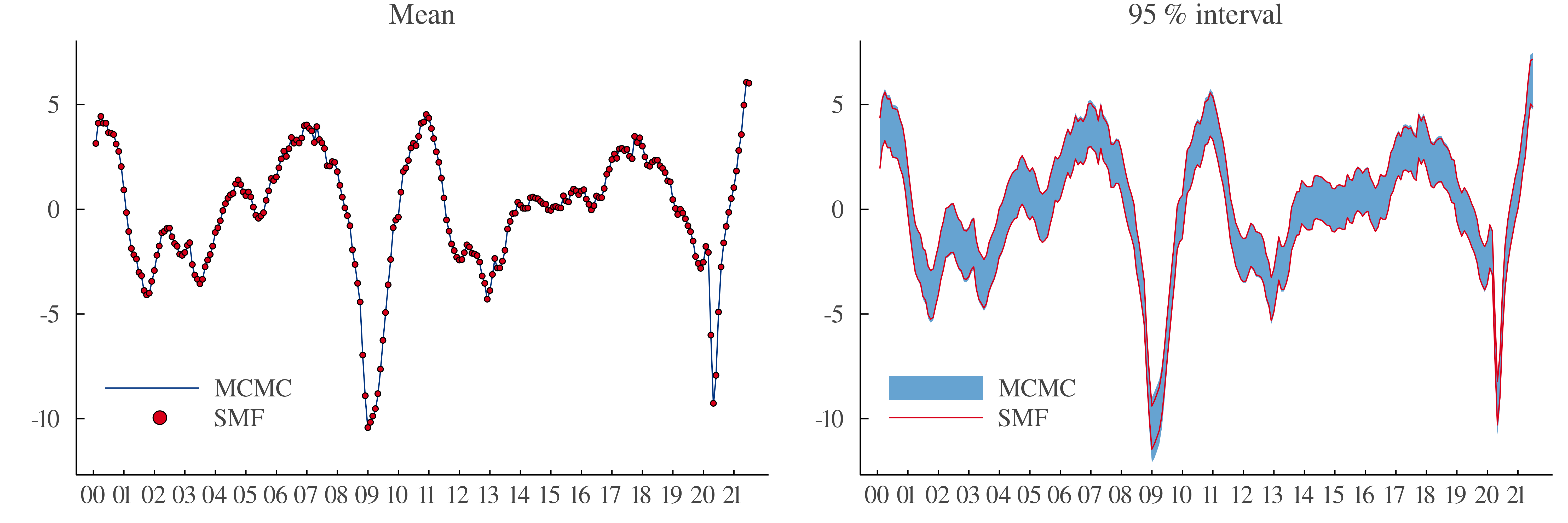}
    \label{sDFMfactor}
\end{figure}

First, we study the approximation for a small DFM of 25 variables and one factor. Figure \ref{sDFMfactor} shows the posterior mean and 95 \% interval of the factor for both MCMC and SMF-approximation. Approximated factor means are very close to the benchmark. The intervals are generally quite close, albeit SMF is a somewhat narrower than MCMC, visually noticeable during the 2008-2009 financial crisis. Table \ref{sDFM_PME1} and \ref{sDFM_PME2} compare errors to the posterior mean, with  mean errors (ME), mean absolute errors (MAE) and root mean square errors (RMSE). The differences are very small for every compared element, including each parameter block and factors and both in- and out-of-sample predictions. 

\begin{table}[H]
\centering
\caption{Posterior mean errors for parameters and in-sample predictions, small model} \label{sDFM_PME1}
\begin{tabular}{lccccc}
    	&	 $[\Phi]_{1,1}$ 	&	 $\lambda_{i,1} \; \forall i,j$ 	&	 $\sigma^2_i \; \forall i$ 	&	 $f_{1,t} \; \forall t$ 	&	 $\hat{\te{y}}_{i,t} \; \forall i,t$	\\ \hline
\multicolumn{1}{l|}{ME} 	&	-0.000	&	0.000	&	-0.007	&	-0.000	&	    \multicolumn{1}{c|}{ -0.000}	\\
\multicolumn{1}{l|}{MAE} 	&	0.000	&	0.001	&	0.007	&	0.024	&	    \multicolumn{1}{c|}{ 0.001}	\\
\multicolumn{1}{l|}{RMSE} 	&	0.000	&	0.001	&	0.008	&	0.030	&	    \multicolumn{1}{c|}{ 0.001}	\\ \cline{2-6}
\end{tabular}
\end{table}

\begin{table}[H]
\centering
\caption{Out-of-sample posterior predictive mean errors, small model} \label{sDFM_PME2}
\begin{tabular}{lcccccc}
    	&	 $\hat{\te{y}}_{i,T+1} \; \forall i$	&	 $\hat{\te{y}}_{i,T+2} \; \forall i$	&	 $\hat{\te{y}}_{i,T+3} \; \forall i$	&	 $\hat{\te{y}}_{i,T+4} \; \forall i$	&	 $\hat{\te{y}}_{i,T+5} \; \forall i$	&	 $\hat{\te{y}}_{i,T+6} \; \forall i$	\\ \hline
\multicolumn{1}{l|}{ME} 	&	0.000	&	-0.001	&	-0.000	&	-0.000	&	-0.001	&	-0.002	\\
\multicolumn{1}{l|}{MAE} 	&	0.002	&	0.002	&	0.002	&	0.002	&	0.003	&	0.004	\\
\multicolumn{1}{l|}{RMSE} 	&	0.003	&	0.003	&	0.003	&	0.003	&	0.004	&	0.005	\\ \cline{2-7}
\end{tabular}
\end{table}

We also look at the SMF coverage probabilities of MCMC-draws. That is, we measure the share of MCMC-draws contained in respective SMF-intervals. Table \ref{sDFM_CP1} shows some small tendencies of narrower intervals for SMF when it comes to parameters and factors. These differences are mainly driven by a few outliers. However, when we make predictive distributions by integrating out parameters and factors, differences goes away almost entirely.

In MCMC, tail draws in factor densities generally correspond with opposite tail draws in parameter densities, counteracting each other when making predictions. In SMF on the other hand, parameters and factors are independent. Overall, the resulting predictive densities become very close. This is also true for out-of-sample predictions, as shown in Table \ref{sDFM_CP2}.

\begin{table}[!ht]
\centering
\caption{SMF coverage probability (\%) of MCMC-draws, small model}  \label{sDFM_CP1}
\begin{tabular}{cccccccccc}
	&	\multicolumn{3}{c}{$[\Phi]_{1,1}$} 	&	 \multicolumn{3}{c}{$\lambda_{i,1} \; \forall i$} 	&	\multicolumn{3}{c}{$\sigma^2_i \; \forall i$}	\\ \cline{2-10}											
\multicolumn{1}{c|}{SMF-interval}	&	Mean	&	Median 	&	\multicolumn{1}{c|}{Stdev} 	&	Mean 	&	Median 	&	\multicolumn{1}{c|}{Stdev} 	&	Mean 	&	Median 	&	\multicolumn{1}{c|}{Stdev} 	\\ \hline
\multicolumn{1}{c|}{50 \%} 	&	49.29	&	49.29	&	\multicolumn{1}{c|}{-}	&	44.94	&	48.95	&	\multicolumn{1}{c|}{9.55}	&	47.69	&	49.93	&	\multicolumn{1}{c|}{6.23}	\\ 
\multicolumn{1}{c|}{75 \%} 	&	74.05	&	74.05	&	\multicolumn{1}{c|}{-}	&	68.30	&	73.85	&	\multicolumn{1}{c|}{13.16}	&	72.07	&	74.88	&	\multicolumn{1}{c|}{8.24}	\\ 
\multicolumn{1}{c|}{95 \%}	&	94.52	&	94.52	&	\multicolumn{1}{c|}{-}	&	89.12	&	94.35	&	\multicolumn{1}{c|}{12.83}	&	92.66	&	94.96	&	\multicolumn{1}{c|}{7.07}	\\ \cline{2-10} 
	&		&		&		&		&		&	\multicolumn{1}{c|}{}	&		&		&		\\
	&	\multicolumn{3}{c}{$f_{1,t} \; \forall t$} 	&	\multicolumn{3}{c|}{$\hat{\te{y}}_{i,t} \; \forall i,t$}	&	\multicolumn{3}{c}{}	\\ \cline{2-7}												
\multicolumn{1}{c|}{SMF-interval}	&	Mean 	&	Median 	&	\multicolumn{1}{c|}{Stdev} 	&	Mean 	&	Median 	&	\multicolumn{1}{c|}{Stdev} 	&		&		&		\\ \cline{1-7}
\multicolumn{1}{c|}{50 \%} 	&	47.26	&	47.72	&	\multicolumn{1}{c|}{2.87}	&	49.99	&	49.99	&	\multicolumn{1}{c|}{0.19}	&		&		&		\\ 
\multicolumn{1}{c|}{75 \%} 	&	71.76	&	72.41	&	\multicolumn{1}{c|}{3.55}	&	74.98	&	74.99	&	\multicolumn{1}{c|}{0.17}	&		&		&		\\ 
\multicolumn{1}{c|}{95 \%}	&	93.07	&	93.60	&	\multicolumn{1}{c|}{2.50}	&	94.98	&	94.99	&	\multicolumn{1}{c|}{0.09}	&		&		&		\\ \cline{2-7} 
\end{tabular}
\end{table}

\begin{table}[H]
\centering
\caption{SMF coverage probability (\%) of MCMC-draws for out-of-sample predictions, small model} \label{sDFM_CP2}
\begin{tabular}{cccccccccc|}
	&	\multicolumn{3}{c}{$\hat{\te{y}}_{i,T+1} \; \forall i$} 	&	\multicolumn{3}{c}{$\hat{\te{y}}_{i,T+2} \; \forall i$} 	&	\multicolumn{3}{c}{$\hat{\te{y}}_{i,T+3} \; \forall i$} 	\\ \cline{2-10}												
\multicolumn{1}{c|}{SMF-interval}	&	Mean	&	Median 	&	\multicolumn{1}{c|}{Stdev} 	&	Mean 	&	Median 	&	\multicolumn{1}{c|}{Stdev} 	&	Mean 	&	Median 	&	\multicolumn{1}{c|}{Stdev} 	\\ \hline
\multicolumn{1}{c|}{50 \%} 	&	49.97	&	49.99	&	\multicolumn{1}{c|}{0.15}	&	50.02	&	50.04	&	\multicolumn{1}{c|}{0.15}	&	50.08	&	50.09	&	\multicolumn{1}{c|}{0.13}	\\ 
\multicolumn{1}{c|}{75 \%} 	&	74.97	&	74.99	&	\multicolumn{1}{c|}{0.12}	&	75.00	&	75.01	&	\multicolumn{1}{c|}{0.12}	&	75.02	&	75.03	&	\multicolumn{1}{c|}{0.16}	\\ 
\multicolumn{1}{c|}{95 \%}	&	95.00	&	95.00	&	\multicolumn{1}{c|}{0.09}	&	95.00	&	95.00	&	\multicolumn{1}{c|}{0.06}	&	94.98	&	94.98	&	\multicolumn{1}{c|}{0.07}	\\ \cline{2-10} 
	&		&		&		&		&		&		&		&		&	\multicolumn{1}{c|}{}	\\
	&	\multicolumn{3}{c}{$\hat{\te{y}}_{i,T+4} \; \forall i$} 	&	\multicolumn{3}{c}{$\hat{\te{y}}_{i,T+5} \; \forall i$} 	&	\multicolumn{3}{c|}{$\hat{\te{y}}_{i,T+6} \; \forall i$} 	\\ \cline{2-10}												
\multicolumn{1}{c|}{SMF-interval}	&	Mean 	&	Median 	&	\multicolumn{1}{c|}{Stdev} 	&	Mean 	&	Median 	&	\multicolumn{1}{c|}{Stdev} 	&	Mean 	&	Median 	&	\multicolumn{1}{c|}{Stdev} 	\\ \hline
\multicolumn{1}{c|}{50 \%} 	&	50.00	&	49.98	&	\multicolumn{1}{c|}{0.17}	&	49.95	&	49.93	&	\multicolumn{1}{c|}{0.17}	&	50.00	&	49.98	&	\multicolumn{1}{c|}{0.20}	\\ 
\multicolumn{1}{c|}{75 \%} 	&	74.95	&	74.96	&	\multicolumn{1}{c|}{0.14}	&	75.00	&	75.00	&	\multicolumn{1}{c|}{0.16}	&	75.03	&	75.02	&	\multicolumn{1}{c|}{0.15}	\\ 
\multicolumn{1}{c|}{95 \%}	&	94.96	&	94.95	&	\multicolumn{1}{c|}{0.08}	&	94.98	&	94.99	&	\multicolumn{1}{c|}{0.07}	&	95.00	&	94.98	&	\multicolumn{1}{c|}{0.08}	\\ \cline{2-10} 
\end{tabular}
\end{table}

\begin{figure}[H]
    \centering
     \caption{Randomly selected parameters and predictions, MCMC-posterior and SMF-approximation, small DFM}
    \includegraphics[width=1\textwidth]{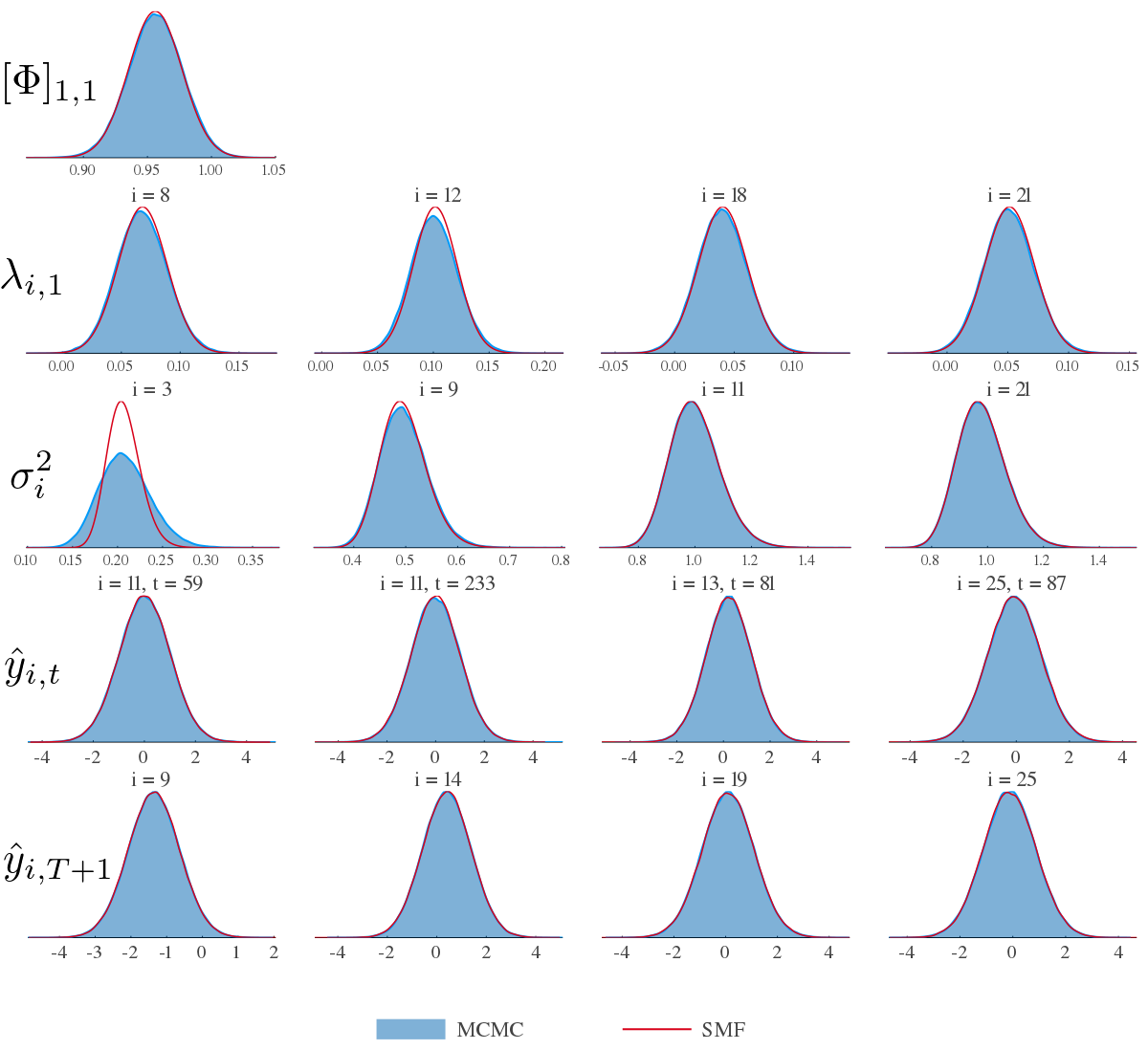}
    \label{sDFMdens}
\end{figure}

Figure \ref{sDFMdens} shows comparing densities for randomly selected elements, which constitute a typical subset. The approximations are again very close. Among them are one outlier parameter with somewhat narrower SMF-density, whereas the rest shows almost indistinguishable approximations. More randomly selected densities are shown in Appendix \ref{sec.C}.

Conclusively, the approximations are close, and very close for predictions. They are obtained in a fraction of computational time, with MCMC taking about 17.5 minutes and SMF less than 0.5 seconds, on a standard laptop \footnote{Intel Core i5-8265U CPU @ 1.60GHz and 8GB RAM}, using the program software Julia, v1.5.4\footnote{See website https://julialang.org/}. SMF converged in 89 iterations\footnote{Convergence critera chosen as  the relative difference in ELBO given in \eqref{criteria}, with tolerance level $10^{-7}$.}, each taking on average about the same computational time as 1.03 MCMC-draws. Thus, SMF was faster than 92 average MCMC-draws. Furthermore, to store approximation results take up much less disk space, as they are fully summarized by density parameters, whereas MCMC-results need to stored as a large number of simulated draws. 


\subsection{Large DFM}
\begin{figure}[H]
    \centering
      \caption[]{\tabular[t]{@{}l@{}} Large DFM factors, MCMC-posterior and SMF-approximation \\ \\\endtabular}
    \includegraphics[width=1\textwidth]{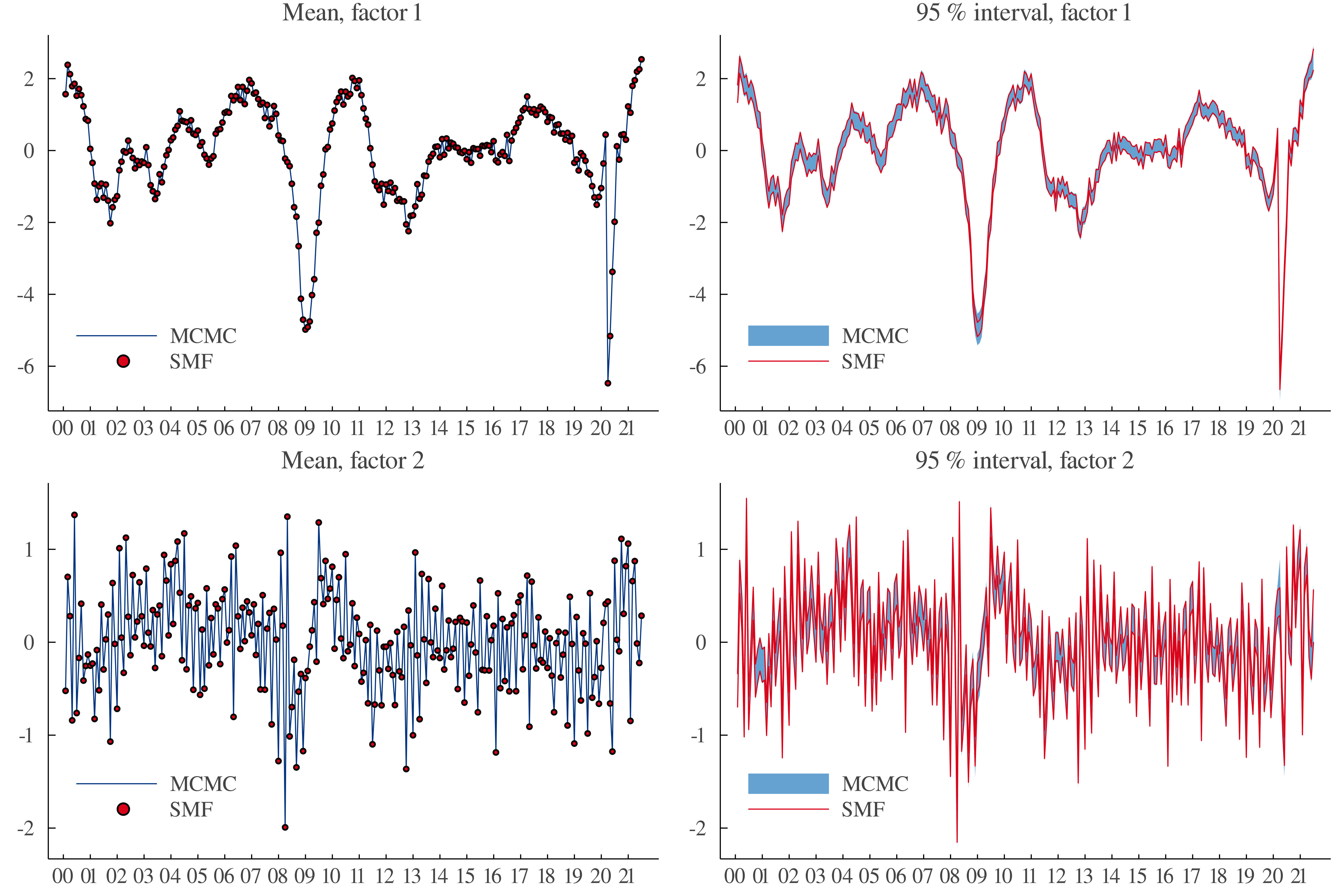}
    \label{lDFMfactor}
\end{figure}

We evaluate the approximation precision for a large DFM with 250 variables, two factors and lags $p=2$. The approximative capabilities are very similar to the small DFM case. Figure \ref{lDFMfactor} shows that SMF captures both factors well. This includes distinct outliers like the initial covid19-crisis in april 2020. Table \ref{lDFM_PME1} and \ref{lDFM_PME2} show very small differences in posterior means to the benchmark. Table \ref{lDFM_CP1} shows SMF coverage probabilities of MCMC-draws, again indicating somewhat narrower variational densities for parameters and factors, and very precise approximations for in-sample predictions. Out-of-sample predicitve densities approximations are also precise, as seen in Table \ref{lDFM_CP2}. 

Densities corresponding to some randomly selected elements are shown in Figure \ref{lDFMdens}, displaying approximate closeness. More randomly selected densities are given in Appendix \ref{sec.C}.

SMF is again a lot faster than MCMC, converging in ELBO in 266 iterations and less than 20 seconds, whereas the MCMC took over 2 hours and 41 minutes. One SMF-iteration took on average the same computational time as about 1.55 MCMC-draws, making the algorithm faster than 413 average MCMC-draws. 

\begin{table}[H]
\centering
\caption{Posterior mean errors for parameters and in-sample predictions, large model} \label{lDFM_PME1}
\begin{tabular}{lccccc}
    	&	 $[\Phi]_{k,j} \forall k,j$ 	&	 $\lambda_{i,j} \; \forall i,j$ 	&	 $\sigma^2_i \; \forall i$ 	&	 $f_{j,t} \; \forall j,t$ 	&	 $\hat{\te{y}}_{i,t} \; \forall i,t$	\\ \hline
\multicolumn{1}{l|}{ME} 	&	-0.000	&	-0.000	&	-0.008	&	-0.000	&	    \multicolumn{1}{c|}{ 0.000}	\\
\multicolumn{1}{l|}{MAE} 	&	0.001	&	0.001	&	0.008	&	0.002	&	    \multicolumn{1}{c|}{ 0.002}	\\
\multicolumn{1}{l|}{RMSE} 	&	0.002	&	0.002	&	0.009	&	0.004	&	    \multicolumn{1}{c|}{ 0.003}	\\ \cline{2-6}
\end{tabular}
\end{table}

\begin{table}[H]
\centering
\caption{Out-of-sample posterior predictive mean errors, small model} \label{lDFM_PME2}
\begin{tabular}{lcccccc}
    	&	 $\hat{\te{y}}_{i,T+1} \; \forall i$	&	 $\hat{\te{y}}_{i,T+2} \; \forall i$	&	 $\hat{\te{y}}_{i,T+3} \; \forall i$	&	 $\hat{\te{y}}_{i,T+4} \; \forall i$	&	 $\hat{\te{y}}_{i,T+5} \; \forall i$	&	 $\hat{\te{y}}_{i,T+6} \; \forall i$	\\ \hline
\multicolumn{1}{l|}{ME} 	&	0.001	&	-0.001	&	0.001	&	-0.003	&	-0.002	&	-0.002	\\
\multicolumn{1}{l|}{MAE} 	&	0.003	&	0.002	&	0.005	&	0.005	&	0.006	&	0.004	\\
\multicolumn{1}{l|}{RMSE} 	&	0.003	&	0.003	&	0.007	&	0.007	&	0.007	&	0.005	\\ \cline{2-7}
\end{tabular}
\end{table}

\begin{figure}[H]
    \centering
     \caption{Randomly selected parameters and predictions, MCMC-posterior and SMF-approximation, large DFM}
    \includegraphics[width=1\textwidth]{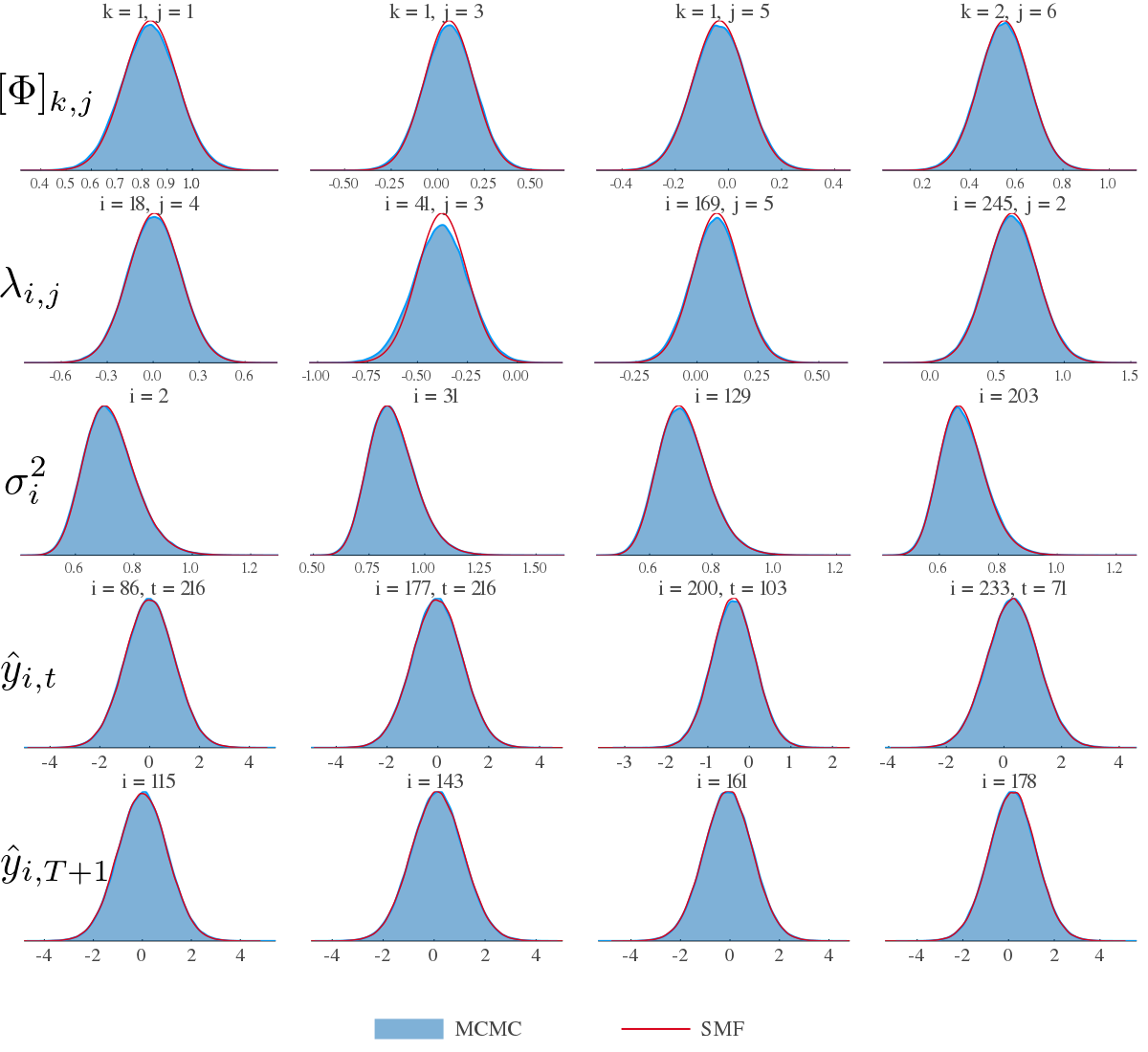}
    \label{lDFMdens}
\end{figure}
\newpage 
\begin{table}[H]
\centering
\caption{SMF coverage probability (\%) of MCMC-draws, large model}  \label{lDFM_CP1}
\begin{tabular}{cccccccccc}
	&	\multicolumn{3}{c}{$[\Phi]_{k,j} \forall k, j$} 	&	 \multicolumn{3}{c}{$\lambda_{i,j} \; \forall i,j$} 	&	\multicolumn{3}{c}{$\sigma^2_i \; \forall i$}	\\ \cline{2-10}												
\multicolumn{1}{c|}{SMF-interval}	&	Mean	&	Median 	&	\multicolumn{1}{c|}{Stdev} 	&	Mean 	&	Median 	&	\multicolumn{1}{c|}{Stdev} 	&	Mean 	&	Median 	&	\multicolumn{1}{c|}{Stdev} 	\\ \hline
\multicolumn{1}{c|}{50 \%} 	&	48.66	&	48.72	&	\multicolumn{1}{c|}{0.63}	&	48.01	&	48.50	&	\multicolumn{1}{c|}{1.79}	&	49.34	&	49.74	&	\multicolumn{1}{c|}{1.04}	\\ 
\multicolumn{1}{c|}{75 \%} 	&	73.53	&	73.54	&	\multicolumn{1}{c|}{0.73}	&	72.72	&	73.30	&	\multicolumn{1}{c|}{2.15}	&	74.26	&	74.71	&	\multicolumn{1}{c|}{1.21}	\\ 
\multicolumn{1}{c|}{95 \%}	&	94.20	&	94.26	&	\multicolumn{1}{c|}{0.41}	&	93.76	&	94.13	&	\multicolumn{1}{c|}{1.40}	&	94.62	&	94.85	&	\multicolumn{1}{c|}{0.67}	\\ \cline{2-10} 
	&		&		&		&		&		&	\multicolumn{1}{c|}{}	&		&		&		\\
	&	\multicolumn{3}{c}{$f_{k,t} \; \forall k,t$} 	&	\multicolumn{3}{c|}{$\hat{\te{y}}_{i,t} \; \forall i,t$}	&	\multicolumn{3}{c}{}	\\ \cline{2-7}												
\multicolumn{1}{c|}{SMF-interval}	&	Mean 	&	Median 	&	\multicolumn{1}{c|}{Stdev} 	&	Mean 	&	Median 	&	\multicolumn{1}{c|}{Stdev} 	&		&		&		\\ \cline{1-7}
\multicolumn{1}{c|}{50 \%} 	&	43.38	&	44.88	&	\multicolumn{1}{c|}{5.82}	&	49.96	&	49.97	&	\multicolumn{1}{c|}{0.19}	&		&		&		\\ 
\multicolumn{1}{c|}{75 \%} 	&	66.92	&	69.11	&	\multicolumn{1}{c|}{8.34}	&	74.96	&	74.96	&	\multicolumn{1}{c|}{0.17}	&		&		&		\\ 
\multicolumn{1}{c|}{95 \%}	&	89.45	&	91.64	&	\multicolumn{1}{c|}{9.53}	&	94.98	&	94.98	&	\multicolumn{1}{c|}{0.09}	&		&		&		\\ \cline{2-7} 
\end{tabular}
\end{table}

\begin{table}[H]
\centering
\caption{SMF coverage probability (\%) of MCMC-draws for out-of-sample predictions, large model} \label{lDFM_CP2}
\begin{tabular}{cccccccccc|}
	&	\multicolumn{3}{c}{$\hat{\te{y}}_{i,T+1} \; \forall i$} 	&	\multicolumn{3}{c}{$\hat{\te{y}}_{i,T+2} \; \forall i$} 	&	\multicolumn{3}{c}{$\hat{\te{y}}_{i,T+3} \; \forall i$} 	\\ \cline{2-10}												
\multicolumn{1}{c|}{SMF-interval}	&	Mean	&	Median 	&	\multicolumn{1}{c|}{Stdev} 	&	Mean 	&	Median 	&	\multicolumn{1}{c|}{Stdev} 	&	Mean 	&	Median 	&	\multicolumn{1}{c|}{Stdev} 	\\ \hline
\multicolumn{1}{c|}{50 \%} 	&	49.92	&	49.93	&	\multicolumn{1}{c|}{0.19}	&	49.94	&	49.94	&	\multicolumn{1}{c|}{0.18}	&	49.96	&	49.97	&	\multicolumn{1}{c|}{0.18}	\\ 
\multicolumn{1}{c|}{75 \%} 	&	74.89	&	74.90	&	\multicolumn{1}{c|}{0.19}	&	74.91	&	74.93	&	\multicolumn{1}{c|}{0.16}	&	74.94	&	74.94	&	\multicolumn{1}{c|}{0.17}	\\ 
\multicolumn{1}{c|}{95 \%}	&	94.94	&	94.95	&	\multicolumn{1}{c|}{0.10}	&	94.95	&	94.95	&	\multicolumn{1}{c|}{0.08}	&	94.96	&	94.96	&	\multicolumn{1}{c|}{0.09}	\\ \cline{2-10} 
	&		&		&		&		&		&		&		&		&	\multicolumn{1}{c|}{}	\\
	&	\multicolumn{3}{c}{$\hat{\te{y}}_{i,T+4} \; \forall i$} 	&	\multicolumn{3}{c}{$\hat{\te{y}}_{i,T+5} \; \forall i$} 	&	\multicolumn{3}{c|}{$\hat{\te{y}}_{i,T+6} \; \forall i$} 	\\ \cline{2-10}												
\multicolumn{1}{c|}{SMF-interval}	&	Mean 	&	Median 	&	\multicolumn{1}{c|}{Stdev} 	&	Mean 	&	Median 	&	\multicolumn{1}{c|}{Stdev} 	&	Mean 	&	Median 	&	\multicolumn{1}{c|}{Stdev} 	\\ \hline
\multicolumn{1}{c|}{50 \%} 	&	49.96	&	49.95	&	\multicolumn{1}{c|}{0.17}	&	49.95	&	49.95	&	\multicolumn{1}{c|}{0.16}	&	49.96	&	49.96	&	\multicolumn{1}{c|}{0.17}	\\ 
\multicolumn{1}{c|}{75 \%} 	&	74.94	&	74.95	&	\multicolumn{1}{c|}{0.15}	&	74.93	&	74.94	&	\multicolumn{1}{c|}{0.15}	&	74.96	&	74.95	&	\multicolumn{1}{c|}{0.15}	\\ 
\multicolumn{1}{c|}{95 \%}	&	94.97	&	94.97	&	\multicolumn{1}{c|}{0.08}	&	94.95	&	94.95	&	\multicolumn{1}{c|}{0.08}	&	94.96	&	94.96	&	\multicolumn{1}{c|}{0.08}	\\ \cline{2-10} 
\end{tabular}
\end{table}

\section{Conclusions}
In this article, we develop a variational inference procedure for estimating dynamic factor models, allowing for arbitrary missing data, including, but not limited to, ragged edge, different sample sizes and mixed frequencies. The aim is to make dynamic factor models more practically viable in fast-pace forecasting settings, while still considering parameter uncertainty and not resorting to point-estimates. We consider a variational density with blockwise independence restrictions between latent factors and parameters. An algorithm is developed minimizing the Kullback-Leibler divergence between the variational density and the true posterior. Our procedure does not require simulation techniques, making it many times faster than ordinary MCMC-methods, potentially reducing hourly long estimations to seconds or a few minutes. 

The estimation algorithm constitutes of iterating a couple of steps: one estimating parameter densities with expressions close to Bayesian linear regression, and one estimating factor densities using Kalman filter and smoother with an augmented observation vector. In empirical examples with Swedish macroeconomic data we find that our variational densities approximate the posterior well, not least predictive densities, which are almost indistinguishable from MCMC-benchmarks.

The computational relief obtained by variational inference opens doors for more intricate and sophisticated specifications and uses for dynamic factor models. We believe there are many potential benefits in building upon our procedure. Future research could delve into  time-varying factor loadings, stochastic volatility, non-gaussian assumptions, hyperparameter selection, sparse solutions and structural analysis. We can leap over some computational hurdles, potentially making dynamic factor models more versatile and reliable tools for real forecasting on a daily basis. 

\bibliography{references}

\newpage
\begin{appendices}
\setcounter{equation}{0}
\renewcommand{\theequation}{\thesection.\arabic{equation}}

\section{VI derivations}\label{sec.A}
This section derives variational densities  and related important object, including ELBO. The variational densities are found by taking the expectation of the log-joint density \eqref{jointdens}, with respect to particular subsets of parameters or latent states.  

\subsection{Likelihood function}\label{sec.A1}

This section follows the derivations of \cite{Spanberg2021}. $Y_A$ and $Y_M$ are available and missing data of the set $Y = \{\ti{y}_t, t=1, ..., T\}$, respectively. \eqref{yeqSS} describes $Y$ as a stochastic process with Gaussian errors. Due to the diagonal property of the covariance matrix $\Sigma_\epsilon$, the conditional density of the would-be-complete data, $p(Y|\theta, F)$, is a joint density of $T \times n$ independent Gaussian variables. The likelihood function is given by integrating out missing data according to: 
\begin{align*}
    p\left(Y_A|\theta,F\right) &= \int p\left(Y|\theta,F\right) dY_M = \int \prod_{i=1}^n\prod_{t=1}^T\left(\frac{1}{\sqrt{2\pi\sigma^2_{i}}}\exp\left\{-\frac{1}{2\sigma_i^2}\left(\ti{y}_{i,t} - \lambda_i' F_t\right)^2 \right\} \right) dY_M. 
\end{align*}
We factorize the available and missing data
\begin{align*}
   p\left(Y_A|\theta,F\right) & = \prod_{i=1}^n\prod_{t=1}^T\left(\frac{1}{\sqrt{2\pi\sigma^2_{i}}}\exp\left\{-\frac{1}{2\sigma_i^2}\left(\ti{y}_{i,t} - \lambda_i' F_t\right)^2 \right\} \right)^{\mathlarger{a_{i,t}}} \\
   &\quad \quad \times \int  \left(\frac{1}{\sqrt{2\pi\sigma^2_{i}}}\exp\left\{-\frac{1}{2\sigma_i^2}\left(\ti{y}_{i,t} - \lambda_i' F_t\right)^2 \right\} \right)^{1-\mathlarger{a_{i,t}}} d\ti{y}_{i,t}.
\end{align*}
As the integral over missing data is Gaussian, and thereby equal to one, we get
\begin{align*}
     p\left(Y_A|\theta,F\right) &= \prod_{i=1}^n\left(\left(\frac{1}{2\pi\sigma_i^2}\right)^{T_i/2} \exp\left\{-\sum_{t=1}^T \frac{a_{i,t}}{2\sigma_i^2}\left(\ti{y}_{i,t} - \lambda_i' F_t\right)^2 \right\} \right),
\end{align*}

where \begin{align*}
    a_{i,t} = \begin{cases} 1 \quad \text{ if } \ti{y}_{i,t} \text{ is available} \\
    0 \quad \text{ if } \ti{y}_{i,t} \text{ is missing} \end{cases}, \quad T_i = \sum_{t=1}^T a_{i,t}.
\end{align*}

In Appendix \ref{sec.A4} we will also make use of the likelihood function in an alternative notation:
\begin{align}
    p\left(Y_A|\theta,F\right) 
    =  C_{Y_A} \exp\left\{-\frac{1}{2}\sum_{t=1}^T\left(\ti{y}_t - \Lambda F_t\right)'A_t\Sigma_\epsilon^{-1}\left(\ti{y}_t - \Lambda F_t\right) \right\}, \label{likeli2}
\end{align}
where $C_{Y_A}$ is constant in terms of $Y_A$ and $A_t = \text{diag}\left(a_{1,t}, a_{2,t}, ..., a_{n,t} \right)$. 

\subsection{Variational density of $\Lambda$ and $\Sigma_\epsilon$} \label{sec.A2}
Log variational density of $\{\Lambda, \Sigma_\epsilon\}$ is given by 
\begin{align*}
    \ln q\left(\Lambda, \Sigma_\epsilon \right) &= \E{q(F)}{\ln p(Y_A, F, \theta)} \\
    &=  C_{\Lambda,{\Sigma_\epsilon}} - \frac{1}{2}\sum_{i=1}^n T_i \ln \sigma_i^2 - \sum_{i=1}^n\sum_{t=1}^T  \frac{a_{i,t}}{2\sigma_i^2}\E{q\left(F\right)}{\left(\ti{y}_{i,t} - \lambda_i' F_t\right)^2} \\
   &\quad -\frac{s}{2}\sum_{i=1}^n \ln \sigma_i^2  -\frac{1}{2} \sum_{i=1}^n \frac{1}{\sigma^2_i} \lambda_i'V^{-1}\lambda_i - \sum_{i=1}^n \left(1 + \nu_i/2\right) \ln \sigma^2_i - \sum_{i=1}^n \frac{\nu_i \tau^2_i}{2\sigma^2_i}, 
\end{align*}
and where $C_{\Lambda, {\Sigma_\epsilon}}$ is constant in terms of $\Lambda$ and $\Sigma_\epsilon$. By expanding square in the expectation we get
\begin{align*}
   \ln q\left(\Lambda, \Sigma_\epsilon \right) &=   C_{\Lambda,{\Sigma_\epsilon}} - \frac{1}{2}\sum_{i=1}^n T_i \ln \sigma_i^2 - \sum_{i=1}^n\sum_{t=1}^T \frac{a_{i,t}}{2\sigma_i^2}\left(\ti{y}_{i,t}^2 - 2\lambda_i' \E{q(F)}{F_t}\ti{y}_{i,t} + \lambda_i' \E{q(F)}{F_t F_t'} \lambda_i \right) \\
   &\quad -\frac{s}{2}\sum_{i=1}^n \ln \sigma_i^2  -\frac{1}{2} \sum_{i=1}^n \frac{1}{\sigma^2_i} \lambda_i'V^{-1}\lambda_i - \sum_{i=1}^n \left(1 + \nu_i/2\right) \ln \sigma^2_i - \sum_{i=1}^n \frac{\nu_i \tau^2_i}{2\sigma^2_i}.
\end{align*}
By rearranging terms (e.g. gather all $\lambda_i$-terms) we have 
\begin{align*}
    \ln q\left(\Lambda, \Sigma_\epsilon \right) &=  C_{\Lambda,{\Sigma_\epsilon}}  -  \sum_{i=1}^n \frac{1}{2\sigma_i^2}\left(-2  \lambda_i'\sum_{t=1}^T \E{q(F)}{F_t}a_{i,t}\ti{y}_{i,t} + \lambda_i'\left[\sum_{t=1}^T a_{i,t}\E{q(F)}{F_t F_t'} + V^{-1}\right] \lambda_i \right) \\
   &\quad -\frac{s}{2}\sum_{i=1}^n \ln \sigma_i^2  - \sum_{i=1}^n \left(1 + \frac{\nu_i + T_i}{2}\right) \ln \sigma^2_i - \sum_{i=1}^n \frac{\nu_i \tau^2_i + \sum_{t=1}^T a_{i,t}\ti{y}_{i,t}^2}{2\sigma^2_i}.
\end{align*}
Completing the square by adding and subtracting the term \begin{align*}  \sum_{i=1}^n \frac{1}{2 \sigma^2_i} \left[\sum_{t=1}^T \E{q(F)}{F_t}a_{i,t}\ti{y}_{i,t}\right]'\left[\sum_{t=1}^T a_{i,t}\E{q(F)}{F_t F_t'} + V^{-1}\right]^{-1}\left[\sum_{t=1}^T \E{q(F)}{F_t}a_{i,t}\te{y}_{i,t}\right] \end{align*}
yields
\begin{align*}
   \ln q\left(\Lambda, \Sigma_\epsilon \right) &=   C_{\Lambda,{\Sigma_\epsilon}}  -  \sum_{i=1}^n \frac{1}{2\sigma_i^2}\left(\lambda_i -  \mu_{\lambda_i}\right)'\Sigma_{\lambda_i}^{-1}\left(\lambda_i -  \mu_{\lambda_i}\right) - \frac{s}{2}\sum_{i=1}^n \ln \sigma_i^2\\
    &\quad - \sum_{i=1}^n \left(1 + \frac{\nu_i + T_i}{2}\right) \ln \sigma^2_i - \sum_{i=1}^n \frac{\nu_i \tau^2_i + \sum_{t=1}^T a_{i,t}\ti{y}_{i,t}^2 - \mu_{\lambda_i}'\Sigma_{\lambda_i}^{-1}\mu_{\lambda_i}}{2\sigma^2_i} \\
    &= \sum_{i=1}^n \ln \mathcal{N}\left(\lambda_i\Big|\mu_{\lambda_i}, \sigma^2_i \Sigma_{\lambda_i} \right) +  \sum_{i=1}^n\ln \text{Scaled Inv-}\chi^2\left(\sigma^2_i\Big|\nu_{\sigma_i}, \tau^2_{\sigma_i} \right), 
\end{align*}
where 
\begin{align*}
       \mu_{\lambda_i} &= \left[\sum_{t=1}^T a_{i,t}\E{q(F)}{F_t F_t'} + V^{-1}\right]^{-1}\left[\sum_{t=1}^T \E{q(F)}{F_t}a_{i,t}\ti{y}_{i,t}\right], \\
       \Sigma_{\lambda_i} &= \left[\sum_{t=1}^T a_{i,t}\E{q(F)}{F_t F_t'} + V^{-1}\right]^{-1}, \\
       \nu_{\sigma_i} &= \nu_i + T_i, \\
       \tau^2_{\sigma_i} &= \frac{1}{\nu_{\sigma_i}}\left(\nu_i\tau^2_i + \sum_{t=1}^T a_{i,t}\ti{y}_{i,t}^2 - \mu_{\lambda_i}'\Sigma_{\lambda_i}^{-1}\mu_{\lambda_i}\right).
\end{align*}
\subsection{Variational density of $\Phi$} \label{sec.A3}
Log variational density of $\Phi$ is given by 
\begin{align*}
    \ln q\left(\Phi \right) &= \E{q(F)}{\ln p(Y_A, F, \theta)} = C_\Phi  -\frac{1}{2}\sum_{t=1}^T\E{q\left(F\right)}{\left(F_t - \widetilde{\Phi} F_{t-1}\right)'\left(F_t - \widetilde{\Phi}F_{t-1}\right)} - \frac{1}{2}\Tr\left(\Phi W^{-1}\Phi'\right) \\
     &=  C_\Phi  -\frac{1}{2}\sum_{t=1}^T\E{q\left(F\right)}{\left(f_t - \Phi F_{t-1}\right)'\left(f_t - \Phi F_{t-1}\right)} - \frac{1}{2}\Tr\left(\Phi W^{-1}\Phi'\right).
\end{align*}
Using the fact that a trace of a scalar is that scalar, cyclical property of traces and that traces commute with the expectation operator gives
\begin{align*}
    \ln q\left(\Phi \right) &=  C_\Phi  -\frac{1}{2}\sum_{t=1}^T \Tr\left(\E{q\left(F\right)}{\left(f_t - \Phi F_{t-1}\right)\left(f_t - \Phi F_{t-1}\right)'}\right) - \frac{1}{2}\Tr\left(\Phi W^{-1}\Phi'\right). 
\end{align*}
Expanding square, collecting $\Phi$-terms and dropping constant term $-\frac{1}{2}\sum_{t=1}^T,  \Tr\left(\E{q\left(F\right)}{f_t f_t'}\right)$, leads to
\begin{align*}
&\ln q\left(\Phi \right) = \\
   &\; C^{(1)}_\Phi - \frac{1}{2}\Tr\left(- \Phi\sum_{t=1}^T\E{q\left(F\right)}{F_{t-1} f_t'} -\sum_{t=1}^T\E{q\left(F\right)}{f_tF_{t-1}'}\Phi' + \Phi\left(\sum_{t=1}^T \E{q\left(F\right)}{F_{t-1} F_{t-1}'} + W^{-1}\right)\Phi'\right). 
\end{align*}
Completing square by adding and subtracting the term
\begin{align*}
    \frac{1}{2}\Tr \left(\left(\sum_{t=1}^T \E{q\left(F\right)}{f_tF_{t-1}'}\right)\left(\sum_{t=1}^T \E{q\left(F\right)}{F_{t-1} F_{t-1}'} + W^{-1}\right)^{-1}\left(\sum_{t=1}^T \E{q\left(F\right)}{F_{t-1}f_t'}\right) \right)
\end{align*}
yields
\begin{align*}
   \ln q\left(\Phi \right)  &= C^{(2)}_\Phi - \frac{1}{2}\Tr\left(\left(\Phi - M_\Phi\right)\Sigma_\Phi^{-1}\left(\Phi - M_\Phi\right)'\right) = \ln \mathcal{MN}\left(M_\Phi, I_r, \Sigma_\Phi\right),
\end{align*}
where
\begin{align*}
    M_\Phi &=  \left(\sum_{t=1}^T \E{q\left(F\right)}{f_tF_{t-1}'}\right)\left(\sum_{t=1}^T \E{q\left(F\right)}{F_{t-1} F_{t-1}'} + W^{-1}\right)^{-1}, \\ 
    \Sigma_\Phi &= \left(\sum_{t=1}^T \E{q\left(F\right)}{F_{t-1} F_{t-1}'} + W^{-1}\right)^{-1}, 
\end{align*}

and where $C_\Phi$, $C_\Phi^{(1)}$ and $C_\Phi^{(2)}$ are constant in terms of $\Phi$.  

\subsection{Variational density of $F$} \label{sec.A4}
Log variational density of $F$ is given by
\begin{align*}
    \ln q\left(F\right) = \E{q(\theta)}{\ln p(Y_A, F, \theta)}
    &= C_F -\frac{1}{2}\sum_{t=1}^T\E{q\left(\theta\right)}{\left(\ti{y}_t - \Lambda F_t\right)'A_t \Sigma_\epsilon^{-1} \left(\ti{y}_t - \Lambda F_t\right)} \\
    &\quad - \frac{1}{2}F_0'\Sigma_{F_0}^{-1}F_0 -\frac{1}{2}\sum_{t=1}^T\E{q\left(\theta\right)}{\left(F_t - \widetilde{\Phi} F_{t-1}\right)'\left(F_t - \widetilde{\Phi} F_{t-1}\right)}, \numberthis \label{q_F1}
\end{align*}
where we have made use of the likelihood notation \eqref{likeli2}. Expanding squares yields
\begin{align*}
    \ln q\left(F\right) &= C_F -\frac{1}{2}\sum_{t=1}^T\left(\te{y}_t'A_t\E{q\left(\theta\right)}{\Sigma_\epsilon^{-1}}\ti{y}_t -  2F_t'\E{q\left(\theta\right)}{\Lambda'A_t\Sigma_\epsilon^{-1}}\ti{y}_t + F_t'\E{q\left(\theta\right)}{\Lambda'A_t\Sigma_\epsilon^{-1}\Lambda} F_t\right) \\
    &\; - \frac{1}{2}F_0'\Sigma_{F_0}^{-1}F_0 -\frac{1}{2}\sum_{t=1}^T\left(F_t'F_t - F_t'\E{q\left(\theta\right)}{\widetilde{\Phi}} F_{t-1} - F_{t-1}'\E{q\left(\theta\right)}{\widetilde{\Phi}'} F_t + F_{t-1}'\E{q\left(\theta\right)}{\widetilde{\Phi}'\widetilde{\Phi}}F_{t-1}\right). \numberthis \label{q_F2}
\end{align*}
Solving \eqref{q_F2} comes down to solving a number of expected values: \begin{itemize}
\item the inverse covariance $\E{q\left(\theta\right)}{\Sigma_\epsilon^{-1}}$,
\item the linear $\{\Lambda, \Sigma_\epsilon\}$-term $\E{q\left(\theta\right)}{\Lambda'A_t\Sigma_\epsilon^{-1}}$,
\item the quadratic $\{\Lambda, \Sigma_\epsilon\}$-term $\E{q\left(\theta\right)}{\Lambda'A_t\Sigma_\epsilon^{-1}\Lambda}$,
\item the linear $\widetilde{\Phi}$-term $\E{q\left(\theta\right)}{\widetilde{\Phi}}$,
\item and the quadratic $\widetilde{\Phi}$-term $\E{q\left(\theta\right)}{\widetilde{\Phi}'\widetilde{\Phi}}$
\end{itemize}  

First we turn to the $\{\Lambda, \Sigma_\epsilon\}$-terms. By properties of distributions in \eqref{lambdasigmadens}, we have $\E{q(\Lambda|\Sigma_\epsilon)}{\Lambda} = M_\Lambda$ and $\E{q(\Sigma_\epsilon)}{\Sigma_\epsilon^{-1}} = \Psi^{-1}$. The linear term is given by iterative expectation:
\begin{align}
    \E{q\left(\theta\right)}{\Lambda'A_t\Sigma_\epsilon^{-1}} = \E{q\left(\Sigma_\epsilon\right)}{\E{q\left(\Lambda|\Sigma_\epsilon\right)}{\Lambda'}A_t\Sigma_\epsilon^{-1}} = M_\Lambda'A_t\Psi^{-1}. \label{linearELambda}
\end{align}

To evaluate the  quadratic term we note that 
\begin{align}
\Lambda'A_t\Sigma_\epsilon^{-1}\Lambda &= \begin{bmatrix} \sum_{i=1}^n \frac{a_{i,t}}{\sigma^2_i} \lambda^2_{i,1} & \sum_{i=1}^n \frac{a_{i,t}}{\sigma^2_i} \lambda_{i,1} \lambda_{i,2} & \dots & \sum_{i=1}^n \frac{a_{i,t}}{\sigma^2_i} \lambda_{i,1} \lambda_{i,n} \\
\sum_{i=1}^n \frac{a_{i,t}}{\sigma^2_i} \lambda_{i,2} \lambda_{i,1} & \sum_{i=1}^n \frac{a_{i,t}}{\sigma^2_i} \lambda^2_{i,2} & \dots & \sum_{i=1}^n \frac{a_{i,t}}{\sigma^2_i} \lambda_{i,2} \lambda_{i,n} \\
\vdots & \vdots & \ddots & \vdots \\ 
\sum_{i=1}^n \frac{a_{i,t}}{\sigma^2_i} \lambda_{i,n} \lambda_{i,1} & \sum_{i=1}^n \frac{a_{i,t}}{\sigma^2_i} \lambda_{i,n} \lambda_{i,2} & \dots & \sum_{i=1}^n \frac{a_{i,t}}{\sigma^2_i} \lambda^2_{i,n}\end{bmatrix} \label{LambdaSquare}
\end{align}

and 
\begin{align}
    \E{q\left(\Lambda|\Sigma_\epsilon\right)}{\lambda_{i,j}\lambda_{g,z}} = m_{i,j}m_{g,z} + \text{Cov}_{q_j\left(\Lambda|\Sigma_\epsilon\right)}\left(\lambda_{i,j},\lambda_{g,z}\right),  \label{Elambdas2}
\end{align}
where $m_{i,j}$ is the $(i,j)$-element of $M_\Lambda$. By the property of distributions \eqref{lambdasigmadens} we have
\begin{align}
    \text{Cov}_{q\left(\Lambda|\Sigma_\epsilon\right)}\left(\lambda_{i,j},\lambda_{g,z}\right) = \begin{cases} \sigma^2_i \left[\Sigma_{\lambda_i}\right]_{j,z} & \text{if } i = g \\
    0 & \text{if } i \neq g \end{cases}. \label{covlambdas}
\end{align} 

Using \eqref{LambdaSquare}-\eqref{covlambdas} we get
\begin{align*} 
&\E{q\left(\Lambda|\Sigma_\epsilon\right)}{\Lambda'A_t\Sigma_\epsilon^{-1}\Lambda} \\
&= \begin{bmatrix} \sum_{i=1}^n \frac{a_{i,t}}{\sigma^2_i} m^2_{i,1} + \sum_{i=1}^n a_{i,t} \left[\Sigma_{\lambda_i}\right]_{1,1} & \dots & \sum_{i=1}^n \frac{a_{i,t}}{\sigma^2_i} m_{i,1} m_{i,n} + \sum_{i=1}^n a_{i,t} \left[\Sigma_{\lambda_i}\right]_{1,n} \\
\vdots & \ddots & \vdots \\ 
\sum_{i=1}^n \frac{a_{i,t}}{\sigma^2_i} m_{i,n} m_{i,1} + \sum_{i=1}^n a_{i,t} \left[\Sigma_{\lambda_i}\right]_{n,1} & \dots & \sum_{i=1}^n \frac{a_{i,t}}{\sigma^2_i} m^2_{i,n} + \sum_{i=1}^n a_{i,t} \left[\Sigma_{\lambda_i}\right]_{n,n}\end{bmatrix} \\ 
&= {M_\Lambda}'A_t \Sigma^{-1}_\epsilon {M_\Lambda} + \sum_{i=1}^n a_{i,t} \Sigma_{\lambda_i}
\end{align*} 
and by iterative expectation
\begin{align}
    \E{q\left(\theta \right)}{\Lambda'A_t\Sigma_\epsilon^{-1}\Lambda} = M_\Lambda'A_t\E{q\left(\Sigma_\epsilon\right)}{\Sigma_\epsilon^{-1}}M_\Lambda + \sum_{i=1}^n a_{i,t} \Sigma_{\lambda_i} = M_\Lambda'A_t\Psi^{-1}M_\Lambda + \sum_{i=1}^n a_{i,t} \Sigma_{\lambda_i}. \label{quadraticELambda}
\end{align} 

Now we turn to $\widetilde{\Phi}$-terms. By properties of distribution \eqref{phidens} we have

\begin{align} 
     \E{q\left(\theta\right)}{\widetilde{\Phi}} \equiv \widetilde{M}_{\Phi} =  \left[
    \begin{array}{cc}
      \multicolumn{2}{c}{M_\Phi}  \\ \hdashline[2pt/2pt]
       \multicolumn{1}{c}{\underset{rp \times rp}{I}} & \multicolumn{1}{;{2pt/2pt}c}{\underset{rp \times r}{0}}
    \end{array}\right] \label{linearEPhi}
\end{align}
and
\begin{align*}
    \E{q\left(\theta\right)}{\Phi'\Phi} = M_\Phi'M_\Phi + \Tr\left(I_r\right)\Sigma_\Phi = M_\Phi'M_\Phi + r\Sigma_\Phi.
\end{align*}
Also note that 
\begin{align*}
    \widetilde{M}_\Phi'\widetilde{M}_\Phi = M_\Phi'M_\Phi + \begin{bmatrix} I_{rp} & 0_{rp \times r} \\ 0_{r \times rp} & 0_{r \times r}\end{bmatrix}.
\end{align*}
Consequently we get
\begin{align}
    \E{q\left(\theta\right)}{\widetilde{\Phi}'\widetilde{\Phi}} = \E{q\left(\theta\right)}{\Phi'\Phi} + \begin{bmatrix} I_{rp} & 0_{rp \times r} \\ 0_{r \times rp} & 0_{r \times r}\end{bmatrix} = \widetilde{M}_\Phi'\widetilde{M}_\Phi + r\Sigma_\Phi. \label{quadraticEPhi}
\end{align}

Inserting expectations \eqref{linearELambda} and \eqref{quadraticELambda}-\eqref{quadraticEPhi} in log density \eqref{q_F2} yields
\begin{align*}
    \ln q(F) &= C_F -\frac{1}{2}\sum_{t=1}^T\left(\ti{y}_t'A_t\Psi^{-1}\ti{y}_t -  2F_t'M_\Lambda'A_t\Psi^{-1}\ti{y}_t + F_t'\left[M_\Lambda'A_t\Psi^{-1}M_\Lambda + \sum_{i=1}^n a_{i,t}\Sigma_{\lambda_i}\right]F_t\right) \\
    &\quad - \frac{1}{2}F_0'\Sigma_{F_0}^{-1}F_0 -\frac{1}{2}\sum_{t=1}^T\left(F_t'F_t - F_t'\widetilde{M}_\Phi F_{t-1} - F_{t-1}'\widetilde{M}_\Phi' F_t + F_{t-1}'\left[\widetilde{M}_\Phi'\widetilde{M}_\Phi + r\Sigma_\Phi\right]F_{t-1}\right). 
\end{align*}
Move parameter covariances $\sum_{i=1}^n a_{i,t}\Sigma_{\lambda_i}$ and $r\Sigma_\Phi$ out of the parentheses according to
\begin{align*}
     \ln q(F) &= C_F -\frac{1}{2}\sum_{t=1}^T\left(\ti{y}_t'A_t\Psi^{-1}\ti{y}_t -  2F_t'M_\Lambda' A_t \Psi^{-1} \ti{y}_t + F_t'M_\Lambda' A_t \Psi^{-1} M_\Lambda F_t\right) \\
    &\quad -\frac{1}{2}\sum_{t=1}^T\left(F_t'F_t - F_t'\widetilde{M}_\Phi F_{t-1} - F_{t-1}'{\widetilde{M}_\Phi}' F_t + F_{t-1}'{\widetilde{M}_\Phi}'\widetilde{M}_\Phi F_{t-1}\right) \\ 
    &\quad- \frac{1}{2}F_0'\Sigma_{F_0}^{-1}F_0  - \frac{1}{2}\sum_{t=0}^{T-1} F_t'r\Sigma_\phi F_t - \frac{1}{2}\sum_{t=1}^T F_t'\left(\sum_{i=1}^n a_{i,t}\Sigma_{\lambda_i}\right)F_t,
\end{align*}
completing squares and rearrange remaining terms after time index
\begin{align*}
    \ln q(F) &= C_F -\frac{1}{2}\sum_{t=1}^T\left(\ti{y}_t - M_\Lambda F_t\right)'A_t \Psi^{-1}\left(\ti{y}_t - M_\Lambda F_t\right) -\frac{1}{2}\sum_{t=1}^T \left(F_t - \widetilde{M}_\Phi F_{t-1}\right)'\left(F_t - \widetilde{M}_\Phi F_{t-1}\right) \\ 
    &\quad- \frac{1}{2}F_0'\left(\Sigma_{F_0}^{-1} + r\Sigma_\Phi\right)F_0  - \frac{1}{2}\sum_{t=1}^{T-1}F_t'\left(\sum_{i=1}^n a_{i,t} \Sigma_{\lambda_i} + r\Sigma_\Phi \right)F_t - \frac{1}{2}F_T'\left(\sum_{i=1}^n a_{i,t}\Sigma_{\lambda_i}\right)F_T
\end{align*}
and inserting the last two terms (parameter covariance terms) in the first square by augmenting vectors and matrices:
\begin{align*}
    \ln q(F) &= C_F -\frac{1}{2}\sum_{t=1}^{T-1}\left(\begin{bmatrix} \ti{y}_t \\ 0_{s \times 1} \end{bmatrix} - \begin{bmatrix} M_\Lambda\\ I_{s} \end{bmatrix}  F_t\right)'\begin{bmatrix}A_t\Psi^{-1} & 0_{n \times s} \\ 0_{s \times n} & \sum_{i=1}^n a_{i,t} \Sigma_{\lambda_i} + r\Sigma_\Phi\end{bmatrix}\left(\begin{bmatrix} \ti{y}_t \\ 0_{s \times 1} \end{bmatrix} - \begin{bmatrix} M_\Lambda \\ I_s \end{bmatrix}  F_t\right)\\
    &\quad -\frac{1}{2}\left(\begin{bmatrix} \ti{y}_T \\ 0_{s \times 1} \end{bmatrix} - \begin{bmatrix} M_\Lambda \\ I_{s} \end{bmatrix}  F_T\right)'\begin{bmatrix}A_T\Psi^{-1}& 0_{n \times s} \\ 0_{s \times n} & \sum_{i=1}^n a_{i,t}\Sigma_{\lambda_i} \end{bmatrix}\left(\begin{bmatrix} \ti{y}_T \\ 0_{s \times 1} \end{bmatrix} - \begin{bmatrix} M_\Lambda \\ I_{s} \end{bmatrix}  F_T\right)\\ 
    &\quad - \frac{1}{2}\sum_{t=1}^T \left(F_t - \widetilde{M}_\Phi F_{t-1}\right)'\left(F_t - \widetilde{M}_\Phi F_{t-1}\right) -  \frac{1}{2}F_0'\left(\Sigma_{F_0}^{-1} + r\Sigma_\Phi\right)F_0. \numberthis \label{q_F3}
\end{align*}
\eqref{q_F3} describes a particular log joint density of $\left\{\widetilde{Y}_A, F\right\}$, which is normalized by some constant, from state space system
\begin{align}
    \widetilde{\ti{y}}_t &= \widetilde{M}_\Lambda F_t + \widetilde{\epsilon}_t, \quad \widetilde{\epsilon}_t \sim \mathcal{N}\left(0, \widetilde{\Sigma}_t \right), \label{Fss1A} \\ 
    F_t &= \widetilde{M}_\Phi F_{t-1} + \begin{bmatrix} I_r \\ 0_{rp \times r} \end{bmatrix}\te{u}_t, \quad \te{u}_t \sim \mathcal{N}\left(0, I_r\right), \label{Fss2A} \\
    F_0 &\sim \mathcal{N}\left(0, \left(\Sigma_{F_0}^{-1} + r\Sigma_\Phi\right)^{-1}\right), \label{Fss3A}
\end{align}
where
\begin{align*}
    \widetilde{\ti{y}}_t &= \begin{bmatrix} \ti{y}_t \\ 0_{s \times 1} \end{bmatrix}, \quad \widetilde{M}_\Lambda = \begin{bmatrix} M_\Lambda \\ I_s \end{bmatrix}, \quad \widetilde{M}_\Phi = \left[
    \begin{array}{cc}
      \multicolumn{2}{c}{M_\Phi}  \\ \hdashline[2pt/2pt]
       \multicolumn{1}{c}{\underset{rp \times rp}{I}} & \multicolumn{1}{;{2pt/2pt}c}{\underset{rp \times r}{0}}
    \end{array}\right], \quad  \widetilde{\Sigma}_t =  \begin{bmatrix} \Psi & 0_{n \times s} \\ 0_{s \times n} & \left(\Sigma_t^\theta\right)^{-1}  \end{bmatrix}, \\ \\
        \Sigma^\theta_t &= \begin{cases}   \sum_{i=1}^n a_{i,t} \Sigma_{\lambda_i} + r\Sigma_\Phi\quad &\text{ if } t=1, ..., T-1, \\ \\
     \sum_{i=1}^n a_{i,T} \Sigma_{\lambda_i} \quad &\text{ if } t=T, \end{cases}
\end{align*}

and where $\widetilde{Y}_A$ are all available elements in $\{\tilde{\ti{y}}_t: t=1,...T\}$. 

However, $q(F)$ is not a joint density with $\widetilde{Y}$, but conditional of $\widetilde{Y}$. This means that there exists a normalization constant within $C_F$ getting us from a joint density expression to a conditional density expression (see Appendix \ref{sec.A4.2}). $q(F)$ can be found by running  \eqref{Fss1A}-\eqref{Fss3A} through the Kalman filter and smoother.  

\subsubsection{Collapsed state space model} \label{sec.A4.1}

By collapsing the observational vector we can greatly reduce computational time, and in our case simply our expressions. 

Let $S^A_t$ be a selection matrix selecting available elements in $\te{y}_t$, i.e. a matrix of rows from a $(n \times n)$ identity matrix, only corresponding to available elements. Note that ${S^A_t}' S^A_t = A_t$. We define objects: 

\begin{itemize} \item Augmented available observational vector $\ti{y}^S_t = \begin{bmatrix} S^A_t\ti{y}_t \\ 0_{s \times 1} \end{bmatrix}$
\item  Time specific matrix mapping state space to available observation space  $M_{\Lambda,t}^S =  \begin{bmatrix} S^A_t M_\Lambda \\ I_s \end{bmatrix}$ 
\item Time specific available observation covariance matrix $\Sigma_t^S = \begin{bmatrix} S^A_t \Psi {S^A_t}' & 0 \\ 0 & \left(\Sigma_t^\theta \right)^{-1}   \end{bmatrix}$. 
\end{itemize} 

\par Following the procedure by \cite{DurbinKoopman2012}, developed by \cite{JungbackerKoopman2015}, the observational vector is transformed 
\begin{align*}  \begin{bmatrix} \ti{y}^\star_t \\ \ti{y}^\diamond_t \end{bmatrix} = \begin{bmatrix} P^\star_t \\ P^\diamond_t \end{bmatrix} \ti{y}^S_t = \begin{bmatrix} F_t \\ 0 \end{bmatrix} + \begin{bmatrix} \epsilon_t^\star \\ \epsilon_t^\diamond \end{bmatrix}, \quad \begin{cases} \epsilon^\star_t \sim \mathcal{N}\left(0, H^\star_t \right) \\ \epsilon^\diamond_t \sim \mathcal{N}\left(0, H^\diamond_t \right)\end{cases}, \end{align*} where $\ti{y}_t^\star$ is the collapsed observational vector and $\ti{y}^\diamond_t$ is independent of $F$. This is achieved by setting $\ti{y}_t^\star$ as the generalized least square estimate of $F_t$ in a conditional distribution of $F_t$ given $\ti{y}^S_t$. $P^\star_t$ is consequently a projection matrix  
\begin{align*} 
    P^\star_t &= \left({M^S_{\Lambda,t}}'\left(\Sigma_t^S\right)^{-1}M^S_{\Lambda,t}\right)^{-1}{M^S_{\Lambda,t}}'\left(\Sigma_t^S\right)^{-1} \\ 
     &= \left(\begin{bmatrix} {M_\Lambda}'{S^A_t}'  & I_s \end{bmatrix} \begin{bmatrix} S^A_t \Psi^{-1} {S^A_t}' & 0 \\ 0 & \Sigma_t^\theta   \end{bmatrix} \begin{bmatrix} S^A_t M_\Lambda  \\ I_s \end{bmatrix} \right)^{-1}\begin{bmatrix} {M_\Lambda}'{S^A_t}'  & I_s \end{bmatrix} \begin{bmatrix} S^A_t \Psi^{-1} {S^A_t}' & 0 \\ 0 & \Sigma_t^\theta \end{bmatrix}  \\
     &= \left({M_{\Lambda}}'{S_t^A}'S_t^A\Psi^{-1} {S_t^A}'S_t^A M_\Lambda + \Sigma^\theta_t \right)^{-1} \begin{bmatrix} {M_{\Lambda}}'{S_t^A}'S_t^A\Psi^{-1} {S_t^A}' & \Sigma^\theta_t \end{bmatrix} \\
     & =  \left({M_{\Lambda}}'A_t\Psi^{-1} M_\Lambda + \Sigma^\theta_t \right)^{-1} \begin{bmatrix} {M_{\Lambda}}'A_t\Psi^{-1} {S_t^A}' & \Sigma^\theta_t \end{bmatrix},
\end{align*}
where we have used the fact that $\Psi$ is diagonal. The collapsed available observational vector is given by
\begin{align*}
   \ti{y}_t^\star &= P^\star_t \ti{y}^S_t = \left({M_{\Lambda}}'A_t\Psi^{-1} M_\Lambda + \Sigma^\theta_t \right)^{-1} \begin{bmatrix} {M_{\Lambda}}'A_t\Psi^{-1} {S_t^A}' & \Sigma^\theta_t \end{bmatrix} \begin{bmatrix} S^A_t \ti{y}_t \\ 0 \end{bmatrix} \\ 
   &= \left({M_{\Lambda}}'A_t\Psi^{-1} M_\Lambda + \Sigma^\theta_t \right)^{-1}{M_{\Lambda}}'\Psi^{-1}A_t\ti{y}_t.
\end{align*}

The covariance matrix of collapsed residual vector $\epsilon_t^\star$ is given by
\begin{align*}
    H^\star_t &= P^\star_t \Sigma_t^S {P^\star_t}' = \left({M^S_{\Lambda,t}}'\left(\Sigma_t^S\right)^{-1}M^S_{\Lambda,t}\right)^{-1}{M^S_{\Lambda,t}}'\left(\Sigma_t^S\right)^{-1}M^S_{\Lambda,t}\left({M^S_{\Lambda,t}}'\left(\Sigma_t^S\right)^{-1}M^S_{\Lambda,t}\right)^{-1} \\
    &= \left({M^S_{\Lambda,t}}'\left(\Sigma_t^S\right)^{-1}M^S_{\Lambda,t}\right)^{-1} = \left({M_{\Lambda}}'A_t\Psi^{-1} M_\Lambda + \Sigma^\theta_t \right)^{-1}.
\end{align*}

As $\ti{y}^\diamond_t$ contains no information about $F$, it's unnecessary to compute $P^\diamond_t$ and $H_t^\diamond$. The collapsed version of the model is given by
\begin{align*}
    \ti{y}^\star_t &=  F_t + \epsilon^\star_t, \quad \epsilon^\star_t \sim \mathcal{N}\left(0, H_t^\star \right),  \\
    F_t &= \widetilde{M}_\Phi F_{t-1} + \begin{bmatrix} I_r \\ 0_{rp \times r} \end{bmatrix}\ti{u}_t, \quad \ti{u}_t \sim \mathcal{N}\left(0, I_r\right), \\
    F_0 &\sim \mathcal{N}\left(0, \left(\Sigma_{F_0}^{-1} + r\Sigma_\Phi\right)^{-1}\right).
\end{align*}

\subsubsection{Constant in variational density of $F$} \label{sec.A4.2}

We derive the constant $C_F$ from \eqref{Fdens}, which is later used to evaluate ELBO (see Appendix \ref{sec.A5}). \eqref{Fdens} describes all non-constant terms in a joint density $q_F\left(F,\widetilde{Y}_A \right)$ for a linear Gaussian state space model. We know that $q(F)$ is a density conditional on $\widetilde{Y}_A$. Let us define $q(F) \equiv q_F\left(F|\widetilde{Y}_A\right)$, then by Bayes' theorem
\begin{align*}
   q_F\left(F|\widetilde{Y}_A\right) =  \frac{q_F\left(F,\widetilde{Y}_A \right)}{ q_F\left(\widetilde{Y}_A\right)}.
\end{align*}
We consequently know that $C_F$ is the sum of constant terms in $\ln q_F\left(F,\widetilde{Y}_A \right)$ subtracted by $\ell_F\left(\widetilde{Y}_A\right) = \ln q_F\left(\widetilde{Y}_A\right)$, i.e
\begin{align*} 
    C_F &= -\frac{1}{2}\sum_{i=1}^n T_i \ln 2\pi - \frac{1}{2} \sum_{i=1}^n T_i \ln \tau^2_{\sigma_i}  - \frac{Ts}{2} \ln 2\pi - \frac{1}{2} \sum_{t=1}^T \ln \det \left(\Sigma_t^\theta\right) \\
    &\quad- \frac{rT + s}{2} \ln 2\pi 
    - \frac{1}{2} \ln \det\left(\left[\Sigma^{-1}_{F_0} + r\Sigma_\Phi \right]^{-1}\right) - \ell_F\left(\widetilde{Y}_A\right). \numberthis \label{C_F1}
\end{align*}
$\ell_F\left(\widetilde{Y}_A\right)$ is the log-likelihood obtained from running \eqref{Fss1}-\eqref{Fss3} through the Kalman filter (not to be confused with $p\left(Y_A|F,\theta\right)$, the likelihood for the entire model \eqref{yeqSS}-\eqref{feqSS}). 

We recommend the computionally faster collapsed state space model, in which $\ell_F\left(\widetilde{Y}_A\right)$ can be decomposed into different parts, including collapsed observations and remainder observations  \citep[see][section 7.2.7]{DurbinKoopman2012}. This is given by
\begin{align*}
\ell_F\left(\widetilde{Y}\right) = \ell_F\left(Y^\star\right) + \ell_F\left(Y^\diamond\right) + \frac{1}{2} \sum_{t=1}^T \ln \det\left(H^\star_t\right) - \frac{1}{2} \sum_{i=1}^n T_i \ln \tau^2_{\sigma_i} - \frac{1}{2}\ln \det\left(\Sigma_t^\theta\right) \numberthis \label{loglikYtA}
\end{align*} 
where the $\ell_F\left(Y^\star\right)$ is the log-likelihood for the collapsed observations obtained from running \eqref{FssC1}-\eqref{FssC3} through the Kalman filter, and $\ell_F\left(Y^\star\right)$ is the remainder observation part. Following the ordinary log-likelihood expression for a linear gaussian state space model \citep[see][section 7.2.1]{DurbinKoopman2012}, the former is given by: 
\begin{align*}
    \ell_F\left(Y^\star\right) = -\frac{Ts}{2}\ln 2\pi - \frac{1}{2} \sum_{t=1}^T \ln \det\left(G_t\right) - \frac{1}{2}\sum_{t=1}^T {\widehat{\epsilon^\star}}'_t G_t^{-1} \widehat{\epsilon^\star}_t,
\end{align*} 
where 
\begin{align}
\widehat{\epsilon^\star}_t &= \ti{y}_t^\star - \E{q(F)}{F_t|Y^\star_{t-1}},  \label{ehatstar}\\
  G_t &= \text{Cov}_{q(F)}\left[F_t|Y^\star_{t-1}\right] + H^\star_t. \label{Gt}
\end{align} 
$\E{q(F)}{F_t|Y^\star_{t-1}}$ and $\text{Cov}_{q(F)}\left[F_t|Y^\star_{t-1}\right]$ are the one-step-ahead predicted state mean and covariance respectively, obtained from the Kalman filter over \eqref{FssC1}-\eqref{FssC3}. 

The remainder observation part is given by
\begin{align*}
    \ell_{F}\left(Y^\diamond\right) = -\frac{1}{2} \sum_{i=1}^n T_i \ln 2\pi - \frac{1}{2} \sum_{t=1}^T {e_t^S}'{\Sigma_t^S}^{-1} e_t^S, 
\end{align*} 
where 
\begin{align}
    e_t^S = \ti{y}_t^S - M^S_{\Lambda,t}\ti{y}_t^\star.  \label{es}
\end{align}
$\Sigma^S_t$, $M^S_{\Lambda,t}$ and $\ti{y}_t^S$ are the time specific augmented covariance, factor loadings and observation vector respectively, defined in Appendix \ref{sec.A4.1}. 

Lastly, by inserting \eqref{loglikYtA} in \eqref{C_F1} we get
\begin{align*}
    C_{F} &= -\frac{rT + s}{2} \ln 2\pi 
    - \frac{1}{2} \ln \det\left(\left[\Sigma^{-1}_{F_0} + r\Sigma_\Phi \right]^{-1}\right) + \frac{1}{2} \ln \left(\frac{\det\left(G_t\right)}{\det\left(H^\star_t\right)} \right) \\ 
    &\quad + \frac{1}{2}\sum_{t=1}^T {\widehat{\epsilon^\star}}'_t G_t^{-1} \widehat{\epsilon^\star}_t + \frac{1}{2} \sum_{t=1}^T {e_t^S}'{\Sigma_t^S}^{-1} e_t^S \numberthis \label{C_F}
\end{align*} 

\subsection{Evidence lower bound} \label{sec.A5}

In this section we derive ELBO described by \eqref{ELBO}. For a SMF approximation of DFM we have the expression
\begin{align}
    \text{ELBO} = \E{q}{\ln p(Y_A, F, \theta)} - \E{q}{\ln q(F)} - \E{q}{\ln q(\theta)}. \label{ELBO1}
\end{align} 
Let us define
\begin{align*}
    \mathcal{E}_F &= -\frac{1}{2}\sum_{t=1}^T\E{q}{\left(\ti{y}_t - \Lambda F_t\right)'A_t \Sigma_\epsilon^{-1} \left(\ti{y}_t - \Lambda F_t\right)} \\
    &\quad - \frac{1}{2}\E{q}{F_0'\Sigma_{F_0}^{-1}F_0} -\frac{1}{2}\sum_{t=1}^T\E{q}{\left(F_t - \widetilde{\Phi} F_{t-1}\right)'\left(F_t - \widetilde{\Phi} F_{t-1}\right)}.
\end{align*}
Then the expectation of the log joint density \eqref{jointdens} can be written as
\begin{align*}
    \E{q}{\ln p(Y_A, F, \theta)} &= \mathcal{E}_F -\frac{1}{2} \sum_{i=1}^n T_i \ln 2\pi -\frac{1}{2} \sum_{i=1}^n T_i \E{q}{\ln \sigma_i^2} -\frac{1}{2} \ln \det\left(\Sigma_{F_0}\right) -\frac{rT + s}{2} \ln 2\pi \\
    &\quad+ \E{q}{\ln p\left(\theta\right)}, \numberthis \label{Ejoint}
\end{align*}
and the expectation of log variational density \eqref{q_F1} as
\begin{align}
    \E{q}{\ln q\left(F\right)} = C_{F} + \mathcal{E}_F. \label{EqF} 
\end{align}
Inserting \eqref{Ejoint}-\eqref{EqF} in \eqref{ELBO1} gives
\begin{align*}
    \text{ELBO} &= -C_{F} -\frac{1}{2} \sum_{i=1}^n T_i \ln 2\pi -\frac{1}{2} \sum_{i=1}^n T_i \E{q}{\ln \sigma_i^2} -\frac{1}{2} \ln \det\left(\Sigma_{F_0}\right) -\frac{rT + s}{2} \ln 2\pi \\ 
    &\quad - \Big(\E{q}{\ln q(\theta)} - \E{q}{\ln p\left(\theta\right)}\Big). \numberthis \label{ELBO2}
\end{align*}
By property of Scaled-inverse-chi-square distribution we have expectation
\begin{align}
    \E{q}{\ln \sigma_i^2} = \ln \nu_{\sigma_i} \tau^2_{\sigma_i} - \ln 2 - \varphi\left(\frac{\nu_{\sigma_i}}{2}\right). \label{Elnsigma}
\end{align}

We also note that $\E{q}{\ln q(\theta)} - \E{q}{\ln p\left(\theta\right)}$ is the KL-divergence between variational densities and prior densities, defined by
\begin{align*} 
\E{q}{\ln q(\theta)} - \E{q}{\ln p\left(\theta\right)} &= \sum_{i=1}^n\E{q\left(\sigma_i^2\right)}{\text{KL}\left(q\left(\lambda_i|\sigma_i^2\right)\Big|\Big|p\left(\lambda_i|\sigma_i^2\right)\right)} + \sum_{i=1}^n\text{KL}\left(q\left(\sigma_i^2\right)\Big|\Big|p\left(\sigma_i^2\right)\right) \\ 
&\quad + \text{KL}\left(q\left(\Phi\right)\Big|\Big|p\left(\Phi\right)\right), 
\end{align*}
due to independence between parameter blocks $\{\Lambda, \Sigma_\epsilon\}$ and $\Phi$. We have Gaussian KL-divergences \citep{Pardo2005}
\begin{align*}
    \E{q\left(\sigma_i^2\right)}{\text{KL}\left(q\left(\lambda_i|\sigma_i^2\right)\Big|\Big|p\left(\lambda_i|\sigma_i^2\right)\right)} &= -\frac{s}{2} + \frac{1}{2}\Tr\left(V^{-1}\Sigma_{\lambda_i}\right) + \frac{1}{2}{\mu_{\lambda_i}}'V^{-1}\mu_{\lambda_i} +  \frac{1}{2}\ln \left(\frac{\det\left(V\right)}{\det\left(\Sigma_{\lambda_i}\right)}\right), \numberthis \label{KLlambda} \\
    \text{KL}\left(q\left(\Phi\right)\Big|\Big|p\left(\Phi\right)\right)  &= -\frac{rs}{2} + \frac{r}{2}\Tr\left(W^{-1}\Sigma_\Phi\right) + \frac{1}{2}\Tr\left(M_\Phi W^{-1} M_\Phi'\right) \\
    &\quad + \frac{r}{2} \ln \left(\frac{\det\left(W\right)}{\det\left(\Sigma_\Phi\right)}\right), \numberthis \label{KLphi}
\end{align*} 
and Scaled-Inverse-chi-square KL-divergence  \citep{LleraBeckmann2016}
\begin{align*}
    \text{KL}\left(q\left(\sigma_i^2\right)\Big|\Big|p\left(\sigma_i^2\right)\right) &= \frac{T_i}{2} \varphi\left(\frac{\nu_{\sigma_i}}{2}\right) - \ln \Gamma\left(\frac{\nu_{\sigma_{i}}}{2}\right) + \ln \Gamma\left(\frac{\nu_i}{2}\right) + \frac{\nu_i}{2}\ln \nu_{\sigma_i} \tau^2_{\sigma_i} - \frac{\nu_i}{2}\ln \nu_i \tau^2_i \\
    &\quad + \frac{\nu_i\tau^2_i - \nu_{\sigma_i}\tau^2_{\sigma_i}}{2\tau^2_{\sigma_i}}, \numberthis \label{KLsigma}
\end{align*}
where $\varphi\left(\cdot\right)$ is the digamma function, and we have used the fact $T_i = \nu_{\sigma_i} - \nu_i$. 

Now we can write a full expression of ELBO. Insert constant \eqref{C_F}, expectation \eqref{Elnsigma} and KL-divergences \eqref{KLlambda}-\eqref{KLsigma} in \eqref{ELBO2} to get
\begin{align}
    \text{ELBO} = \{F\text{-terms}\} + \{\Lambda\text{-terms}\} + \{\Phi\text{-terms}\} + \{\Sigma_\epsilon\text{-terms}\}
\end{align}
where 
\begin{align*}
    \{F\text{-terms}\} &= -\frac{1}{2}\sum_{i=1}^n T_i \ln \pi - \frac{1}{2}\ln \left(\frac{\det\left(\Sigma_{F_0}\right)}{\det\left(\left[\Sigma_{F_0}^{-1} + r\Sigma_\Phi\right]^{-1}\right)}\right) - \frac{1}{2} \ln \left(\frac{\det\left(G_t\right)}{\det\left(H^\star_t\right)} \right) \\
    &\quad - \frac{1}{2}\sum_{t=1}^T {\widehat{\epsilon^\star}}'_t G_t^{-1} \widehat{\epsilon^\star}_t - \frac{1}{2} \sum_{t=1}^T {e_t^S}'{\Sigma_t^S}^{-1} e_t^S, \\
    \{\Lambda\text{-terms}\} &= \frac{ns}{2} - \frac{1}{2}\sum_{i=1}^n\Tr\left(V^{-1}\Sigma_{\lambda_i}\right) - \frac{1}{2}\sum_{i=1}^n{\mu_{\lambda_i}}'V^{-1}\mu_{\lambda_i} -  \frac{1}{2}\sum_{i=1}^n\ln \left(\frac{\det\left(V\right)}{\det\left(\Sigma_{\lambda_i}\right)}\right), \\
    \{\Phi\text{-terms}\} &= \frac{rs}{2} - \frac{r}{2}\Tr\left(W^{-1}\Sigma_\Phi\right) - \frac{1}{2}\Tr\left(M_\Phi W^{-1} M_\Phi'\right) - \frac{r}{2} \ln \left(\frac{\det\left(W\right)}{\det\left(\Sigma_\Phi\right)}\right), \\
    \{\Sigma_\epsilon\text{-terms}\} &= \sum_{i=1}^n\ln \Gamma\left(\frac{\nu_{\sigma_{i}}}{2}\right) - \sum_{i=1}^n\ln \Gamma\left(\frac{\nu_i}{2}\right) - \sum_{i=1}^n\frac{\nu_{\sigma_i}}{2}\ln \nu_{\sigma_i} \tau^2_{\sigma_i} + \sum_{i=1}^n\frac{\nu_i}{2}\ln \nu_i \tau^2_i + \sum_{i=1}^n\frac{\nu_{\sigma_i}\tau^2_{\sigma_i} - \nu_i\tau^2_i}{2\tau^2_{\sigma_i}}.
\end{align*} 
To aid the reader and any researchers wanting to utilize this result, we compile a reiterated list of necessary objects to evaluate the ELBO expression above: 
\begin{itemize}
\item scalars $r$, $s$, $n$ and $T_i$, $\forall i=1, ..., n$,
\item prior parameters $V$, $W$, $\Sigma_{F_0}$ $\nu_i$, and $\tau_i^2$, $\forall i=1, ..., n$,
\item variational parameters $\mu_{\lambda_i}$,$\Sigma_{\lambda_i}$ $\nu_{\sigma_i}$, and $\tau^2_{\sigma_i}$, $\forall i=1, ..., n$, given by \eqref{mulambda}-\eqref{tausigma}
\item variational parameters $M_\Phi$ and $\Sigma_\Phi$ given by \eqref{M_Phi}-\eqref{Sigma_Phi}
\item collapsed observations $\ti{y}_t^\star$ and respective residual covariance $H_t^\star$, $\forall t=1, ..., T$ given by \eqref{ystar}-\eqref{Hstar}, 
\item one-step-ahead predicted collapsed residuals $\hat{\epsilon^\star}_t$ and respective one-step-ahead prediction covariance $G_t$, $\forall t=1, ..., T$, given by \eqref{ehatstar}-\eqref{Gt}, 
\item remainder residual $\epsilon_t^S$ given by \eqref{es}, aggregated available observation vector $\ti{y}_t^S$ and the corresponding parameter matrices $M_{\Lambda,t}^S$ and $\Sigma_t^S$, $\forall t=1, ..., T$ defined in Appendix \ref{sec.A4.1}
\end{itemize}
\newpage 
\section{Data}\label{sec.B}
\setcounter{table}{0}
\renewcommand{\thetable}{B\arabic{table}}

This section describes the data set used in the empirical examples. The set consists of 250 monthly macroeconomic variables, ranging from 2000m1-2021m06 with varying availability. Data was obtained from the Macrobond database plattform\footnote{See website https://www.macrobond.com/.} on 20th August 2021. All series are standardized with mean 0 and standard deviation 1.   We give information about the data set in in Table \ref{datatable}, including variable descriptions, Macrobond codes, data availability original sources, transformations and adjustments. Table \ref{datacodes} describes the table codes, aiding the interpretation of Table \ref{datatable}.  

\small
\begin{table}[H]
\centering
\caption{Data Table Codes} \label{datacodes}
\begin{tabular}{cl}
\hline
SM	&		\\	
X[$i$]	&	Included in small DFM as variable $i$	\\	\hline
Source	&		\\	
1	&	Statistics Sweden	\\	
2	&	National Institute of Economic Research	\\	
3	&	Central Bank of Sweden	\\	
4	&	Swedbank	\\	
5	&	Skandiabanken	\\	
6	&	Skandinaviska Enskilda Banken AB	\\	
7	&	SBAB	\\	
8	&	Nordea	\\	
9	&	Länsförsäkringar	\\	
10	&	Danske Bank	\\	
11	&	Swedish Public Employment Service	\\	
12	&	Valueguard Sweden AB	\\	
13	&	Svensk Mäklarstatistik	\\	
14	&	Nasdaq OMX Nordic	\\	
15	&	Swedish Transport Agency	\\	\hline
Trans.	&		\\	
1	&	No transformation	\\	
2	&	Log-differencing	\\	
3	&	Differencing 	\\	\hline
SA	&		\\	
N	&	No seasonal adjustment	\\	
S	&	Seasonal adjustment by original source	\\	
A	&	Seasonal adjustment by author (X12-arima)	\\	\hline
\end{tabular}
\end{table}
\newpage
\singlespacing 
\begin{table}[H]
\caption{Data Table} \label{datatable}
\scriptsize
\begin{tabularx}{\textwidth}{p{0.001\textwidth}Xlp{0.001\textwidth}c
ccc}
$i$	&	Variable	&	Macrobond	&	SM	&	Sample	&	Source	&	Trans.	&	SA	\\	\hline
	&	\textbf{National Accounts}	&		&		&		&		&		&		\\	
1	&	GDP-indicator	&	senaac4900	&	X1	&	2011m02-2021m06	&	1	&	2	&	S	\\	\hline
	&	\textbf{Production Value Index}	&		&		&		&		&		&		\\	
2	&	Private Sector Excluding Financial \& Insurance Services	&	seprod2690	&		&	2010m02-2021m06	&	1	&	2	&	S	\\	
3	&	Intermediate Goods	&	seprod2693	&		&	2010m02-2021m06	&	1	&	2	&	S	\\	
4	&	Energy Related Goods Excluding Electricy, Gas, Steam \& Hot Water Plans	&	seprod2694	&		&	2010m02-2021m06	&	1	&	2	&	S	\\	
5	&	Capital Goods	&	seprod2696	&		&	2010m02-2021m06	&	1	&	2	&	S	\\	
6	&	Non-Durable Consumer Goods	&	seprod2697	&		&	2010m02-2021m06	&	1	&	2	&	S	\\	
7	&	Durable Consumer Goods	&	seprod2698	&		&	2010m02-2021m06	&	1	&	2	&	S	\\	
8	&	Mines \& Quarries	&	seprod2699	&		&	2010m02-2021m06	&	1	&	2	&	S	\\	
9	&	Manufacturing	&	seprod2706	&		&	2010m02-2021m06	&	1	&	2	&	S	\\	
10	&	Food Product, Beverage \& Tobacco	&	seprod2707	&		&	2010m02-2021m06	&	1	&	2	&	S	\\	
11	&	Textiles, Clothing \& Leather	&	seprod2711	&		&	2010m02-2021m06	&	1	&	2	&	S	\\	
12	&	Wood \& Products of Wood, Cork, Cane, Etc., Except Furniture	&	seprod2715	&		&	2010m02-2021m06	&	1	&	2	&	S	\\	
13	&	Pulp, Paper \& Paperboard	&	seprod2719	&		&	2010m02-2021m06	&	1	&	2	&	S	\\	
14	&	Printers; Other Industry for Recorded Media	&	seprod2723	&		&	2010m02-2021m06	&	1	&	2	&	S	\\	
15	&	Manufacture of Coke \& Refined Petroleum Products	&	seprod2724	&		&	2010m02-2021m06	&	1	&	2	&	S	\\	
16	&	Chemical, Chemical Products \& Pharmaceutical Products	&	seprod2725	&		&	2010m02-2021m06	&	1	&	2	&	S	\\	
17	&	Rubber \& Plastic Products	&	seprod2728	&		&	2010m02-2021m06	&	1	&	2	&	S	\\	
18	&	Other Non-Metallic Mineral Products	&	seprod2729	&		&	2010m02-2021m06	&	1	&	2	&	S	\\	
19	&	Basic Metals	&	seprod2730	&		&	2010m02-2021m06	&	1	&	2	&	S	\\	
20	&	Fabricated Metal Products, Except Machinery \& Equipment	&	seprod2732	&		&	2010m02-2021m06	&	1	&	2	&	S	\\	
21	&	Computer, Electronic \& Optical Products	&	seprod2734	&		&	2010m02-2021m06	&	1	&	2	&	S	\\	
22	&	Electrical Equipment	&	seprod2735	&		&	2010m02-2021m06	&	1	&	2	&	S	\\	
23	&	Machinery \& Equipment N.E.C	&	seprod2736	&		&	2010m02-2021m06	&	1	&	2	&	S	\\	
24	&	Transport Equipment	&	seprod2737	&		&	2010m02-2021m06	&	1	&	2	&	S	\\	
25	&	Other Manufacturing, Repair \& Installation of Machinery \& Equipment	&	seprod2740	&		&	2010m02-2021m06	&	1	&	2	&	S	\\	
26	&	Electricity, Gas, Steam \& Hot Water Plants	&	seprod2745	&		&	2010m02-2021m06	&	1	&	2	&	S	\\	
27	&	Water Works; Sewage Plants, Waste-Disposal Plants	&	seprod2747	&		&	2010m02-2021m06	&	1	&	2	&	S	\\	
28	&	Construction	&	seprod2748	&		&	2010m02-2021m06	&	1	&	2	&	S	\\	
29	&	Service Sector Excluding Financial \& Insurance Activities	&	seprod2749	&		&	2010m02-2021m06	&	1	&	2	&	S	\\	
30	&	Trade; Repear Establishments for Motor Vehicles \& Motorcycles	&	seprod2751	&		&	2010m02-2021m06	&	1	&	2	&	S	\\	
31	&	Land Transport Companies \& Pipeline Transport Companies	&	seprod2756	&		&	2010m02-2021m03	&	1	&	2	&	S	\\	
32	&	Shipping Companies	&	seprod2757	&		&	2010m02-2021m03	&	1	&	2	&	S	\\	
33	&	Airline Companies	&	seprod2758	&		&	2010m02-2021m03	&	1	&	2	&	S	\\	
34	&	Warehousing Companies \& Service Companies Supporting Transport	&	seprod2759	&		&	2010m02-2021m03	&	1	&	2	&	S	\\	
35	&	Postal \& Courier Companies	&	seprod2760	&		&	2010m02-2021m03	&	1	&	2	&	S	\\	
36	&	Hotels \& Restaurants	&	seprod2761	&		&	2010m02-2021m06	&	1	&	2	&	S	\\	
37	&	Information \& Communication Companies	&	seprod2763	&		&	2010m02-2021m06	&	1	&	2	&	S	\\	
38	&	Real Estate Companies	&	seprod2770	&		&	2010m02-2021m06	&	1	&	2	&	S	\\	
39	&	Professional, Scientific \& Techinical Companies	&	seprod2771	&		&	2010m02-2021m06	&	1	&	2	&	S	\\	
40	&	Education, Human Health \& Social Work Activities	&	seprod2773	&		&	2010m02-2021m06	&	1	&	2	&	S	\\	
41	&	Art, Entertainment, Recreation \& Other Service Activities	&	seprod2776	&		&	2010m02-2021m06	&	1	&	2	&	S	\\	\hline
	&	\textbf{Economic Tendency Survey}	&		&		&		&		&		&		\\	
42	&	Economic Tendency Indicator	&	sesurv0258	&	X2	&	2000m01-2021m06	&	2	&	1	&	N	\\	
43	&	Manufacturing, Confidence Indicator	&	sesurv0256	&		&	2010m05-2021m06	&	2	&	1	&	S	\\	
44	&	Construction, Confidence Indicator	&	sesurv0250	&		&	2000m01-2021m06	&	2	&	1	&	S	\\	
45	&	Retail Trade, Conficence Indicator	&	sesurv0620	&		&	2000m01-2021m06	&	2	&	1	&	S	\\	
46	&	Pricate Service Sector, Confidence Indicator	&	sesurv0622	&		&	2000m01-2021m06	&	2	&	1	&	S	\\	
47	&	The Consumer Confidence Indicator	&	sesurv0270	&		&	2000m01-2021m06	&	2	&	1	&	S	\\	
48	&	The Macro Index	&	sesurv0271	&		&	2000m01-2021m06	&	2	&	1	&	N	\\	
49	&	The Micro Index	&	sesurv0272	&		&	2000m01-2021m06	&	2	&	1	&	N	\\	\hline
	&	\textbf{Purchasing Managers' Index}	&		&		&		&		&		&		\\	
50	&	Total Manufacturing	&	sesurv0177	&	X3	&	2000m01-2021m06	&	4	&	1	&	S	\\	
51	&	Manufacturing, New Orders	&	sesurv0178	&		&	2000m01-2021m06	&	4	&	1	&	S	\\	
52	&	Manufacturing, Production	&	sesurv0179	&		&	2000m01-2021m06	&	4	&	1	&	S	\\	
53	&	Manufacturing, Employment	&	sesurv0180	&		&	2000m01-2021m06	&	4	&	1	&	S	\\	
54	&	Manufacturing, Delivery Times	&	sesurv0181	&		&	2000m01-2021m06	&	4	&	1	&	S	\\	
55	&	Manufacturing, Inventories	&	sesurv0182	&		&	2000m01-2021m06	&	4	&	1	&	S	\\	
56	&	Manufacturing, Export Orders	&	sesurv0183	&		&	2000m01-2021m06	&	4	&	1	&	S	\\	
57	&	Manufacturing, Domestic Orders	&	sesurv0184	&		&	2000m01-2021m06	&	4	&	1	&	S	\\	
58	&	Manufacturing, Back-Log of Orders	&	sesurv0185	&		&	2000m01-2021m06	&	4	&	1	&	S	\\	
59	&	Manufacturing, Prices	&	sesurv0186	&		&	2000m01-2021m06	&	4	&	1	&	S	\\	
60	&	Manufacturing, Import	&	sesurv0187	&		&	2000m01-2021m06	&	4	&	1	&	S	\\	
61	&	Manufacturing, Planned Production	&	sesurv0188	&		&	2000m01-2021m06	&	4	&	1	&	S	\\	
62	&	Total Services	&	sesurv0242	&	X4	&	2005m10-2021m06	&	4	&	1	&	S	\\	
63	&	Services, Business Volume	&	sesurv0243	&		&	2005m10-2021m06	&	4	&	1	&	S	\\	
64	&	Services, Delivery Times	&	sesurv0244	&		&	2005m10-2021m06	&	4	&	1	&	S	\\	
65	&	Serivces, New Orders	&	sesurv0245	&		&	2005m10-2021m06	&	4	&	1	&	S	\\	
66	&	Services, Employment	&	sesurv0246	&		&	2005m10-2021m06	&	4	&	1	&	S	\\	
67	&	Services, Back-Log of Orders	&	sesurv0247	&		&	2005m10-2021m06	&	4	&	1	&	S	\\	
68	&	Services, Commodity \& Intermediate Goods Prices	&	sesurv0248	&		&	2005m10-2021m06	&	4	&	1	&	S	\\	
69	&	Services, Planned Business Volume	&	sesurv0249	&		&	2005m10-2021m06	&	4	&	1	&	S	\\	\hline
	&	\textbf{Interest Rates}	&		&		&		&		&		&		\\	
70	&	Policy Rate, Repo Rate	&	serate0001	&	X5	&	2000m01-2021m06	&	3	&	3	&	N	\\	
71	&	Government Benchmarks, 2 Year, Yield	&	se2yrbgov	&		&	2000m01-2021m06	&	3	&	3	&	N	\\	
72	&	Government Benchmarks, 5 Year, Yield	&	se5yrbgov	&		&	2000m01-2021m06	&	3	&	3	&	N	\\	
73	&	Government Benchmarks, 7 Year, Yield	&	se7yrbfix	&		&	2000m01-2021m06	&	3	&	3	&	N	\\	
74	&	Government Benchmarks, 10 Year, Yield	&	se10yrbfix	&		&	2000m01-2021m06	&	3	&	3	&	N	\\	
75	&	Mortage Lending Rates, 3 Month, Swedbank	&	swedbank3m	&		&	2000m01-2021m06	&	4	&	3	&	N	\\	
76	&	Mortage Lending Rates, 3 Month, Skandiabanken	&	sandiabrr3m	&		&	2013m01-2021m06	&	5	&	3	&	N	\\	
77	&	Mortage Lending Rates, 3 Month, SEB	&	seb3m	&		&	2005m07-2021m06	&	6	&	3	&	N	\\	
78	&	Mortage Lending Rates, 3 Month, SBAB	&	sbab3m	&		&	2000m01-2021m06	&	7	&	3	&	N	\\	
79	&	Mortage Lending Rates, 3 Month, Nordea	&	nordea3m	&		&	2000m01-2021m06	&	8	&	3	&	N	\\	
80	&	Mortage Lending Rates, 3 Month, Länsförsäkringar	&	lfbank3m	&		&	2010m03-2021m06	&	9	&	3	&	N	\\	
81	&	Mortage Lending Rates, 3 Year, Danske Bank	&	ddbk3y	&		&	2015m04-2021m06	&	10	&	3	&	N	\\	\hline
\end{tabularx} \null\hfill (continues)
\end{table}

\scriptsize
\begin{tabularx}{\textwidth}{p{0.001\textwidth}Xlp{0.001\textwidth}c
ccc}
$i$	&	Variable	&	Macrobond	&	SM	&	Sample	&	Source	&	Trans.	&	SA	\\	\hline
	&	\textbf{Unemployment}	&		&		&		&		&		&		\\	
82	&	Unemployment Rate (SCB) 15-74 Year old	&	selama0635	&	X6	&	2001m02-2021m06	&	1	&	3	&	S	\\	
83	&	Unemployment Rate (PES) 16-64 Year old	&	selama2242	&		&	2000m01-2021m06	&	11	&	3	&	S	\\	
84	&	Newly Registered Unemployed, PES	&	selama4466	&		&	2000m01-2021m06	&	11	&	3	&	A	\\	
85	&	Unemployed Persons, Duration 1 Week	&	selama1864	&		&	2005m05-2021m06	&	1	&	2	&	A	\\	
86	&	Unemployed Persons, Duration 2 Weeks	&	selama1879	&		&	$\begin{cases} \text{05m05-20m02}, \\ \text{20m05-20m09}, \\ \text{21m01} \\ \text{21m05-21m06} \end{cases}$	&	1	&	2	&	A	\\	
87	&	Unemployed Persons, Duration 3-4 Weeks	&	selama0572	&		&	2005m05-2021m06	&	1	&	2	&	A	\\	
88	&	Unemployed Persons, Duration 5-26 Weeks	&	selama0587	&		&	2005m05-2021m06	&	1	&	2	&	A	\\	
89	&	Unemployed Persons, Duration over 27 Weeks	&	selama0602	&		&	2005m05-2021m06	&	1	&	2	&	A	\\	
90	&	Unemployed Persons, Average Weeks	&	selama11493	&	X7	&	2005m05-2021m06	&	1	&	3	&	S	\\	\hline
	&	\textbf{Employment}	&		&		&		&		&		&		\\	
91	&	Employed, Total	&	selama2960	&	X8	&	2001m02-2021m06	&	1	&	2	&	S	\\	
92	&	Employed, Domestic Born	&	selama10002	&		&	2005m05-2021m06	&	1	&	2	&	S	\\	
93	&	Employed, Foreign Born	&	selama10014	&		&	2005m05-2021m06	&	1	&	2	&	S	\\	
94	&	Employed, Agriculture, Forestry \& Fishing	&	selama1175	&		&	2009m02-2021m06	&	1	&	2	&	A	\\	
95	&	Employed, Construction	&	selama1178	&		&	2009m02-2021m06	&	1	&	2	&	A	\\	
96	&	Employed, Education	&	selama1185	&		&	2009m02-2021m06	&	1	&	2	&	A	\\	
97	&	Employed, Financial Intermediation, Business Services	&	selama1183	&		&	2009m02-2021m06	&	1	&	2	&	A	\\	
98	&	Employed, Health Care	&	selama1186	&		&	2000m01-2021m06	&	1	&	2	&	A	\\	
99	&	Employed, Hotels \& Restaurants	&	selama1181	&		&	2009m02-2021m06	&	1	&	2	&	A	\\	
100	&	Employed, Information \& Communication	&	selama1182	&		&	2009m02-2021m06	&	1	&	2	&	A	\\	
101	&	Employed, Manufacture of Machinery \& Transport	&	selama1177	&		&	2009m02-2021m06	&	1	&	2	&	A	\\	
102	&	Employed, Manufacturing \& Mining, Energy \& Enviroment	&	selama1176	&		&	2000m01-2021m06	&	1	&	2	&	A	\\	
103	&	Employed, Personal \& Cultural Services	&	selama1187	&		&	2009m02-2021m06	&	1	&	2	&	A	\\	
104	&	Employed, Public Administration etc.	&	selama1184	&		&	2009m02-2021m06	&	1	&	2	&	A	\\	
105	&	Employed, Trade	&	selama1179	&		&	2000m01-2021m06	&	1	&	2	&	A	\\	
106	&	Employed, Transport	&	selama1180	&		&	2009m02-2021m06	&	1	&	2	&	A	\\	
107	&	Employed, 15-24 Year Old, excl. Working Abroad	&	selama2811	&		&	2005m05-2021m06	&	1	&	2	&	A	\\	
108	&	Employed, 25-54 Year Old, excl. Working Abroad	&	selama2812	&		&	2005m05-2021m06	&	1	&	2	&	A	\\	
109	&	Employed, 55-74 Year Old, excl. Working Abroad	&	selama2813	&		&	2005m05-2021m06	&	1	&	2	&	A	\\	
110	&	Employed, Government Sector (16-64 Year Old)	&	selama0332	&		&	2000m01-2021m06	&	1	&	2	&	A	\\	
111	&	Employed, Municipal Sector (16-64 Year Old)	&	selama0337	&		&	2000m01-2021m06	&	1	&	2	&	A	\\	
112	&	Employed, Private Sector (16-64 Year Old)	&	selama0342	&		&	2000m01-2021m06	&	1	&	2	&	A	\\	
113	&	Employed, Working Abroad (16-64 Year Old)	&	selama2820	&		&	2005m05-2021m06	&	1	&	2	&	A	\\	\hline
	&	\textbf{Working Hours per Week}	&		&		&		&		&		&		\\	
114	&	Agriculture, Forestry \& Fishery	&	selama2467	&		&	2009m02-2021m06	&	1	&	2	&	A	\\	
115	&	Manufacturing \& Mining, Energy \& Environment	&	selama2470	&		&	2000m01-2021m06	&	1	&	2	&	A	\\	
116	&	Manufacture of Machinery \& Transport	&	selama2473	&		&	2000m01-2021m06	&	1	&	2	&	A	\\	
117	&	Construction	&	selama2476	&		&	2000m01-2021m06	&	1	&	2	&	A	\\	
118	&	Trade	&	selama2479	&		&	2000m01-2021m06	&	1	&	2	&	A	\\	
119	&	Transport	&	selama2482	&		&	2009m02-2021m06	&	1	&	2	&	A	\\	
120	&	Hotels \& Restaurants	&	selama2485	&		&	2009m02-2021m06	&	1	&	2	&	A	\\	
121	&	Information \& Communication	&	selama2488	&		&	2009m02-2021m06	&	1	&	2	&	A	\\	
122	&	Financial Intermediation, Business Services	&	selama2491	&		&	2009m02-2021m06	&	1	&	2	&	A	\\	
123	&	Public Administration	&	selama2494	&		&	2009m02-2021m06	&	1	&	2	&	A	\\	
124	&	Education	&	selama2497	&		&	2009m02-2021m06	&	1	&	2	&	A	\\	
125	&	Health Care	&	selama2500	&		&	2000m01-2021m06	&	1	&	2	&	A	\\	
126	&	Personal \& Cultural Services	&	selama2503	&		&	2009m02-2021m06	&	1	&	2	&	A	\\	\hline
	&	\textbf{Labor Costs}	&		&		&		&		&		&		\\	
127	&	Salaries, Manufacturing, Mining \& Quarrying, Energy \& Environmental Companies	&	selama1761	&		&	2008m05-2021m05	&	1	&	2	&	A	\\	
128	&	Salaries, Construction	&	selama1768	&		&	2008m05-2021m05	&	1	&	2	&	A	\\	
129	&	Salaries, Wholesale \& Retail Trade; Accomodation \& Food Services	&	selama1769	&		&	2008m05-2021m05	&	1	&	2	&	A	\\	
130	&	Salaries, Information \& Communication	&	selama1774	&		&	2008m05-2021m05	&	1	&	2	&	A	\\	
131	&	Salaries, Financial \& Insurance Activities, Real Estate Activities, Business Services	&	selama1775	&		&	2008m05-2021m05	&	1	&	2	&	A	\\	
132	&	Salaries, Education, Healthcare, Personal \& Cultural Services	&	selama1780	&		&	2008m05-2021m05	&	1	&	2	&	A	\\	
133	&	Wages, Manufacturing, Mining \& Quarrying, Energy \& Environmental Companies	&	selama1786	&		&	2008m05-2021m05	&	1	&	2	&	A	\\	
134	&	Wages, Construction	&	selama1793	&		&	2008m05-2021m05	&	1	&	2	&	A	\\	
135	&	Wages, Wholesale \& Retail Trade; Accomodation \& Food Services	&	selama1794	&		&	2008m05-2021m05	&	1	&	2	&	A	\\	
136	&	Wages, Financial \& Insurance Activities, Real Estate Activities, Business Services	&	selama1799	&		&	2008m05-2021m05	&	1	&	2	&	A	\\	
137	&	Wages, Education, Healthcare, Personal \& Cultural Services	&	selama1802	&		&	2008m05-2021m05	&	1	&	2	&	A	\\	\hline
	&	\textbf{Labor Turnover}	&		&		&		&		&		&		\\	
138	&	Redundancy Notices	&	selama2268	&	X9	&	2000m01-2021m06	&	11	&	1	&	S	\\	
139	&	New Vacancies	&	selama0005	&	X10	&	2000m01-2021m06	&	11	&	2	&	A	\\	\hline
	&	\textbf{Consumer Price Index}	&		&		&		&		&		&		\\	
140	&	Underlying Inflation CPIF	&	sepric1091	&	X11	&	2000m01-2021m06	&	1	&	2	&	A	\\	
141	&	Food \& Non-Alcoholic Beverages	&	sepric0020	&		&	2000m01-2021m06	&	1	&	2	&	A	\\	
142	&	Alcoholic Beverages \& Tobacco	&	sepric0021	&		&	2000m01-2021m06	&	1	&	2	&	A	\\	
143	&	Clothing \& Footwear	&	sepric0022	&		&	2000m01-2021m06	&	1	&	2	&	A	\\	
144	&	Owner-Occupied Housing; Water \& Dwelling Services	&	sepric1060	&		&	2000m01-2021m06	&	1	&	2	&	A	\\	
145	&	Electricity	&	sepric1062	&		&	2000m01-2021m06	&	1	&	2	&	A	\\	
146	&	Fuel	&	sepric1401	&		&	2005m02-2021m06	&	1	&	2	&	A	\\	
147	&	Rented \& Housing Co-Operative Dwellings: Rent incl. Heating	&	sepric1063	&		&	2000m01-2021m06	&	1	&	2	&	A	\\	
148	&	Imputed Rent for Owner Occupied Housing 	&	sepric1064	&		&	2000m01-2021m06	&	1	&	2	&	A	\\	
149	&	Furnishings \& Household Goods	&	sepric0024	&		&	2000m01-2021m06	&	1	&	2	&	A	\\	
150	&	Health Care	&	sepric0025	&		&	2000m01-2021m06	&	1	&	2	&	A	\\	
151	&	Transport	&	sepric0026	&		&	2000m01-2021m06	&	1	&	2	&	A	\\	
152	&	Communication	&	sepric0027	&		&	2000m01-2021m06	&	1	&	2	&	A	\\	
153	&	Recreation \& Culture	&	sepric0028	&		&	2000m01-2021m06	&	1	&	2	&	A	\\
154	&	Education	&	sepric1092	&		&	2000m01-2021m06	&	1	&	2	&	A	\\	
155	&	Restaurants \& Hotels	&	sepric0029	&		&	2000m01-2021m06	&	1	&	2	&	A	\\	
156	&	Miscellaneous Goods \& Services	&	sepric0030	&		&	2000m01-2021m06	&	1	&	2	&	A	\\	\hline
\end{tabularx} \null\hfill (continues)

\scriptsize
\begin{tabularx}{\textwidth}{p{0.001\textwidth}Xlp{0.001\textwidth}c
ccc}
$i$	&	Variable	&	Macrobond	&	SM	&	Sample	&	Source	&	Trans.	&	SA	\\	\hline
	&	\textbf{Producer Price Index}	&		&		&		&		&		&	A	\\	
157	&	Total	&	sepric3001	&	X12	&	2000m01-2021m06	&	1	&	2	&	A	\\	
158	&	Products of Agriculture, Forestry \& Fishing	&	sepric3002	&		&	2000m01-2021m06	&	1	&	2	&	A	\\	
159	&	Mining \& Quarrying	&	sepric3024	&		&	$\begin{cases} \text{00m01-10m12}, \\ \text{13m02-21m06}  \end{cases}$ 	&	1	&	2	&	A	\\	
160	&	Food Products; Beverages	&	sepric3071	&		&	2000m01-2021m06	&	1	&	2	&	A	\\	
161	&	Textiles, Wearing Apparel, Leather \& Leather Products	&	sepric3092	&		&	2000m01-2021m06	&	1	&	2	&	A	\\	
162	&	Wood, Products of Wood \& Cork, Except Furniture; Articles of Straw \& Plaiting Materials	&	sepric3105	&		&	2000m01-2021m06	&	1	&	2	&	A	\\	
163	&	Paper \& Paper Products	&	sepric3115	&		&	2000m01-2021m06	&	1	&	2	&	A	\\	
164	&	Printing \&n Reproduction Services of Recorded Media	&	sepric3126	&		&	$\begin{cases} \text{08m02-16m12} \\ \text{18m02-21m06} \end{cases}$ 	&	1	&	2	&	A	\\	
165	&	Coke \& Refined Petroleum Products	&	sepric3132	&		&	2000m01-2021m06	&	1	&	2	&	A	\\	
166	&	Chemicals \& Chemical Products	&	sepric3135	&		&	2000m01-2021m06	&	1	&	2	&	A	\\	
167	&	Basic Pharmaceutical Products	&	sepric3155	&		&	$\begin{cases} \text{0m01-11m12} \\ \text{13m02-14m12} \\ \text{19m02-21m06} \end{cases}$	&	1	&	2	&	A	\\	
168	&	Rubber \& Plastic Products	&	sepric3157	&		&	2000m01-2021m06	&	1	&	2	&	A	\\	
169	&	Other Non-Metallic Mineral Products	&	sepric3166	&		&	2000m01-2021m06	&	1	&	2	&	A	\\	
170	&	Basic Metals	&	sepric3192	&		&	2000m01-2021m06	&	1	&	2	&	A	\\	
171	&	Fabricated Metal Products, Except Machinery \& Equipment	&	sepric3214	&		&	2000m01-2021m06	&	1	&	2	&	A	\\	
172	&	Computer, Electronic \& Optical Products	&	sepric3259	&		&	2000m01-2021m06	&	1	&	2	&	A	\\	
173	&	Machinery \& Equipment N.E.C	&	sepric3278	&		&	2000m01-2021m06	&	1	&	2	&	A	\\	
174	&	Motor Vehicles, Trailers \& Semi-Trailers; Other Transport Equipment	&	sepric3312	&		&	2000m01-2021m06	&	1	&	2	&	A	\\	
175	&	Furniture, Other Manufactured Goods	&	sepric3329	&		&	2000m01-2021m06	&	1	&	2	&	A	\\	
176	&	Repair \& Installation Services of Machinery \& Equipment	&	sepric3340	&		&	2005m02-2021m06	&	1	&	2	&	A	\\	
177	&	Electricy, Gas, Steam \& Air Conditioning	&	sepric3350	&		&	2000m01-2021m06	&	1	&	2	&	A	\\	
178	&	Water Supply; Sewage, Waist Management \& remediation Services	&	sepric3359	&		&	2000m01-2021m06	&	1	&	2	&	A	\\	
179	&	Export Prices	&	sepric3786	&	X13	&	2000m01-2021m06	&	1	&	2	&	A	\\	
180	&	Import Prices	&	sepric4120	&	X14	&	2000m01-2021m06	&	1	&	2	&	A	\\	\hline
	&	\textbf{Real Estate Prices and Sales}	&		&		&		&		&		&		\\	
181	&	HOX index, residential	&	secons0199	&	X15	&	2005m02-2021m06	&	12	&	2	&	A	\\	
182	&	One- \& Two-Dwelling Buildings, Purchase-Price-Coefficient (K/T)	&	secons0173	&		&	2003m04-2021m06	&	1	&	2	&	A	\\	
183	&	One- \& Two-Dwelling Buildings, Permanent Dwelling, Number of Purchases	&	secons0170	&	X16	&	2000m01-2021m06	&	1	&	2	&	A	\\	
184	&	Houses, Residential, Purchase Price	&	secons2426	&		&	2005m02-2021m06	&	13	&	2	&	A	\\	
185	&	Tenant-Owned Flats, Residential, Purchase Price	&	secons2722	&		&	2005m02-2021m06	&	13	&	2	&	A	\\	
186	&	Houses, Number of Transactions	&	secons2414	&		&	2005m02-2021m06	&	13	&	2	&	A	\\	
187	&	Tenant-Owned Flats, Number of Transactions	&	secons2710	&		&	2005m02-2021m06	&	13	&	2	&	A	\\	
	&	\textbf{Equity Prices}	&		&		&		&		&		&		\\	\hline
188	&	OMX Stockholm Index	&	omxspi	&	X17	&	2000m01-2021m06	&	14	&	2	&	N	\\	
189	&	OMX Stockholm Large Cap, number of trades	&	seeqst0012	&		&	2011m9-2021m06	&	14	&	2	&	N	\\	
190	&	OMX Stockholm Large Cap, turnover velocity	&	seeqst0013	&		&	2011m8-2021m06	&	14	&	2	&	N	\\	
191	&	OMX Stockholm Large Cap, total turnover	&	seeqst0011	&		&	2011m9-2021m06	&	14	&	2	&	N	\\	
192	&	OMX Stockholm Mid Cap, number of trades	&	seeqst0020	&		&	2011m9-2021m06	&	14	&	2	&	N	\\	
193	&	OMX Stockholm Mid Cap, turnover velocity	&	seeqst0021	&		&	2011m8-2021m06	&	14	&	2	&	N	\\	
194	&	OMX Stockholm Mid Cap, total turnover	&	seeqst0019	&		&	2011m9-2021m06	&	14	&	2	&	N	\\	
195	&	OMX Stockholm Small Cap, number of trades	&	seeqst0028	&		&	2011m9-2021m06	&	14	&	2	&	N	\\	
196	&	OMX Stockholm Small Cap, turnover velocity	&	seeqst0029	&		&	2011m8-2021m06	&	14	&	2	&	N	\\	
197	&	OMX Stockholm Small Cap, total turnover	&	seeqst0027	&		&	2011m9-2021m06	&	14	&	2	&	N	\\	\hline
	&	\textbf{Airport Statistics}	&		&		&		&		&		&		\\	
198	&	Freight, Foreign, Total	&	setran0151	&		&	$\begin{cases} \text{07m02-11m11} \\ \text{12m02-18m12} \\ \text{20m02-21m06} \end{cases}$	&	15	&	2	&	A	\\	
199	&	Freight, Domestic, Outgoing	&	setran0153	&		&	$\begin{cases} \text{07m02-11m11} \\ \text{12m02-18m12} \\ \text{20m02-21m06} \end{cases}$	&	15	&	2	&	A	\\	
200	&	Passengers, Foreign, Total	&	setran0932	&		&	2013m02-2021m06	&	15	&	2	&	A	\\	
201	&	Passengers, Domestic, Outgoing	&	setran0933	&		&	2013m02-2021m06	&	15	&	2	&	A	\\	\hline
	&	\textbf{Turism}	&		&		&		&		&		&		\\	
202	&	Number of Nights Spent by Non-Residents, All Accomodation Establishments	&	setour0638	&		&	2008m02-2021m06	&	1	&	2	&	A	\\	
203	&	Number of Nights Spent by Non-Residents, Excluding Metropolitan	&	setour0642	&		&	2008m02-2021m06	&	1	&	2	&	A	\\	
204	&	Nights, Camping	&	setour0154	&		&	2008m02-2021m06	&	1	&	2	&	A	\\	
205	&	Nights, Commercially Arranged Rentals in Private Cottages \& Apartments	&	setour0205	&		&	2008m02-2021m06	&	1	&	2	&	A	\\	
206	&	Nights, Holiday Villages	&	setour0052	&		&	2000m01-2021m06	&	1	&	2	&	A	\\	
207	&	Nights, Hotels	&	setour0001	&		&	2000m01-2021m06	&	1	&	2	&	A	\\	
208	&	Nights, Youth Hostels	&	setour0103	&		&	2000m01-2021m06	&	1	&	2	&	A	\\	\hline
	&	\textbf{Household Consumption Indicator}	&		&		&		&		&		&		\\	
209	&	Household Consumption	&	setrad1516	&	X18	&	2000m02-2021m06	&	1	&	2	&	S	\\	
210	&	Furnitures, Furnishings, Housholds equipment \& Consumables	&	setrad2833	&		&	2000m02-2021m06	&	1	&	2	&	S	\\	
211	&	Housing, Electricity, Gas \& Heating	&	setrad2832	&		&	2000m02-2021m06	&	1	&	2	&	S	\\	
212	&	Other Goods \& Services	&	setrad2838	&		&	2000m02-2021m06	&	1	&	2	&	S	\\	
213	&	Post \& Telecommunications	&	setrad2835	&		&	2000m02-2021m06	&	1	&	2	&	S	\\	
214	&	Recreation \& Culture, Goods \& Services	&	setrad2836	&		&	2000m02-2021m06	&	1	&	2	&	S	\\	
215	&	Restarurants, Cafes, Hotels \& Other Accomodation Services	&	setrad2837	&		&	2000m02-2021m06	&	1	&	2	&	S	\\	
216	&	Retail Trade, Clothing \& Footwear	&	setrad2831	&		&	2000m02-2021m06	&	1	&	2	&	S	\\	
217	&	Retail Trade, Mostly Food \& Beverages	&	setrad2830	&		&	2000m02-2021m06	&	1	&	2	&	S	\\	
218	&	Transport \& Retail Sales \& Services of Motor Vehicles	&	setrad2834	&		&	2000m02-2021m06	&	1	&	2	&	S	\\	\hline
	&	\textbf{Domestic Trade}	&		&		&		&		&		&		\\	
219	&	Retail Trade, Total except Fuel	&	setrad0477	&	X19	&	2000m01-2021m06	&	1	&	2	&	S	\\	
220	&	Retail Trade, Mostly Food Except State Liquor Stores	&	setrad1891	&		&	2000m01-2021m06	&	1	&	2	&	S	\\	
221	&	Mostly Durables	&	setrad0482	&		&	2000m01-2021m06	&	1	&	2	&	S	\\	\hline
\end{tabularx} \null\hfill (continues)

\scriptsize
\begin{tabularx}{\textwidth}{p{0.001\textwidth}Xcp{0.003\textwidth}c
ccc}
$i$	&	Variable	&	Macrobond	&	SM	&	Sample	&	Source	&	Trans.	&	SA	\\	\hline
	&	\textbf{Foreign Trade}	&		&		&		&		&		&		\\	
222	&	Export, Goods	&	setrad0001	&	X20	&	2000m01-2021m06	&	1	&	2	&	A	\\	
223	&	Import, Goods	&	setrad0000	&	X21	&	2000m01-2021m06	&	1	&	2	&	A	\\	
224	&	Export, Food \& Live Animals	&	setrad1503	&		&	2000m01-2021m05	&	1	&	2	&	A	\\	
225	&	Export, Beverages \& Tobacco	&	setrad1504	&		&	2000m01-2021m05	&	1	&	2	&	A	\\	
226	&	Export, Crude Materials, Inedible, except Fuels	&	setrad1505	&		&	2000m01-2021m05	&	1	&	2	&	A	\\	
227	&	Export, Mineral Fuels, Lubricants \& Related Materials	&	setrad1506	&		&	2000m01-2021m05	&	1	&	2	&	A	\\	
228	&	Export, Animal \& Vegetable Oils, Fats \& Waxes	&	setrad1507	&		&	2000m01-2021m05	&	1	&	2	&	A	\\	
229	&	Export, Chemicals \& Related Products	&	setrad1508	&		&	2000m01-2021m05	&	1	&	2	&	A	\\	
230	&	Export, Manufactured Goods Classified by Material	&	setrad1509	&		&	2000m01-2021m05	&	1	&	2	&	A	\\	
231	&	Export, Machinery \& Transport eqipment	&	setrad1510	&		&	2000m01-2021m05	&	1	&	2	&	A	\\	
232	&	Export, Miscellaneous Manufactured Articles	&	setrad1511	&		&	2000m01-2021m05	&	1	&	2	&	A	\\	
233	&	Export, Goods N.E.C	&	setrad1512	&		&	2000m01-2021m05	&	1	&	2	&	A	\\	
234	&	Import, Foor \& Live Animals	&	setrad1493	&		&	2000m01-2021m05	&	1	&	2	&	A	\\	
235	&	Import, Beverages \& Tobacco	&	setrad1494	&		&	2000m01-2021m05	&	1	&	2	&	A	\\	
236	&	Import, Crude Materials, Inedible except Fuels	&	setrad1495	&		&	2000m01-2021m05	&	1	&	2	&	A	\\	
237	&	Import, Mineral Fuels, Lubricants \& Related Materials	&	setrad1496	&		&	2000m01-2021m05	&	1	&	2	&	A	\\	
238	&	Import, Animal \& Vegetable Oils, Fats \& Waxes	&	setrad1497	&		&	2000m01-2021m05	&	1	&	2	&	A	\\	
239	&	Import, Chemicals \& Related Products N.E.S	&	setrad1498	&		&	2000m01-2021m05	&	1	&	2	&	A	\\	
240	&	Import, Manufactured Goods Classfied by Material	&	setrad1499	&		&	2000m01-2021m05	&	1	&	2	&	A	\\	
241	&	Import, Machinery \& Transport Equipment	&	setrad1500	&		&	2000m01-2021m05	&	1	&	2	&	A	\\	
242	&	Import, Miscellaneous Manufactured Articles	&	setrad1501	&		&	2000m01-2021m05	&	1	&	2	&	A	\\	
243	&	Import, Goods N.E.C	&	setrad1502	&		&	2000m01-2021m05	&	1	&	2	&	A	\\	\hline
	&	\textbf{Bankruptcies}	&		&		&		&		&		&		\\	
244	&	All Employees, Incorporated Companies	&	sebank0403	&	X22	&	2000m01-2021m06	&	1	&	3	&	A	\\	
245	&	All Employees, Deceased Estates	&	sebank0404	&		&	2013m02-2021m06	&	1	&	3	&	A	\\	
246	&	All Employees, Sole Proprietorships	&	sebank0405	&		&	2000m01-2021m06	&	1	&	3	&	A	\\	
247	&	All Employees, Partnerships	&	sebank0406	&	X23	&	2000m01-2021m06	&	1	&	3	&	A	\\	
248	&	All Employees, Private Individuals	&	sebank0407	&		&	2000m01-2021m06	&	1	&	3	&	A	\\	\hline
	&	\textbf{Exchange Rates}	&		&		&		&		&		&		\\	
249	&	SEK per EUR	&	eurrbfix	&	X24	&	2000m01-2021m06	&	3	&	2	&	N	\\	
250	&	SEK per USD	&	usdrbfix	&	X25	&	2000m01-2021m06	&	3	&	2	&	N	\\	\hline
\end{tabularx} \null\hfill
\newpage  
\section{Additional Figures} \label{sec.C}
\setcounter{figure}{0}
\renewcommand\thefigure{\thesection.\arabic{figure}}
\begin{figure}[!ht]
    \centering
     \caption{Randomly selected parameters and predictions, MCMC-posterior and SMF-approximation, small DFM}
    \includegraphics[width=1\textwidth]{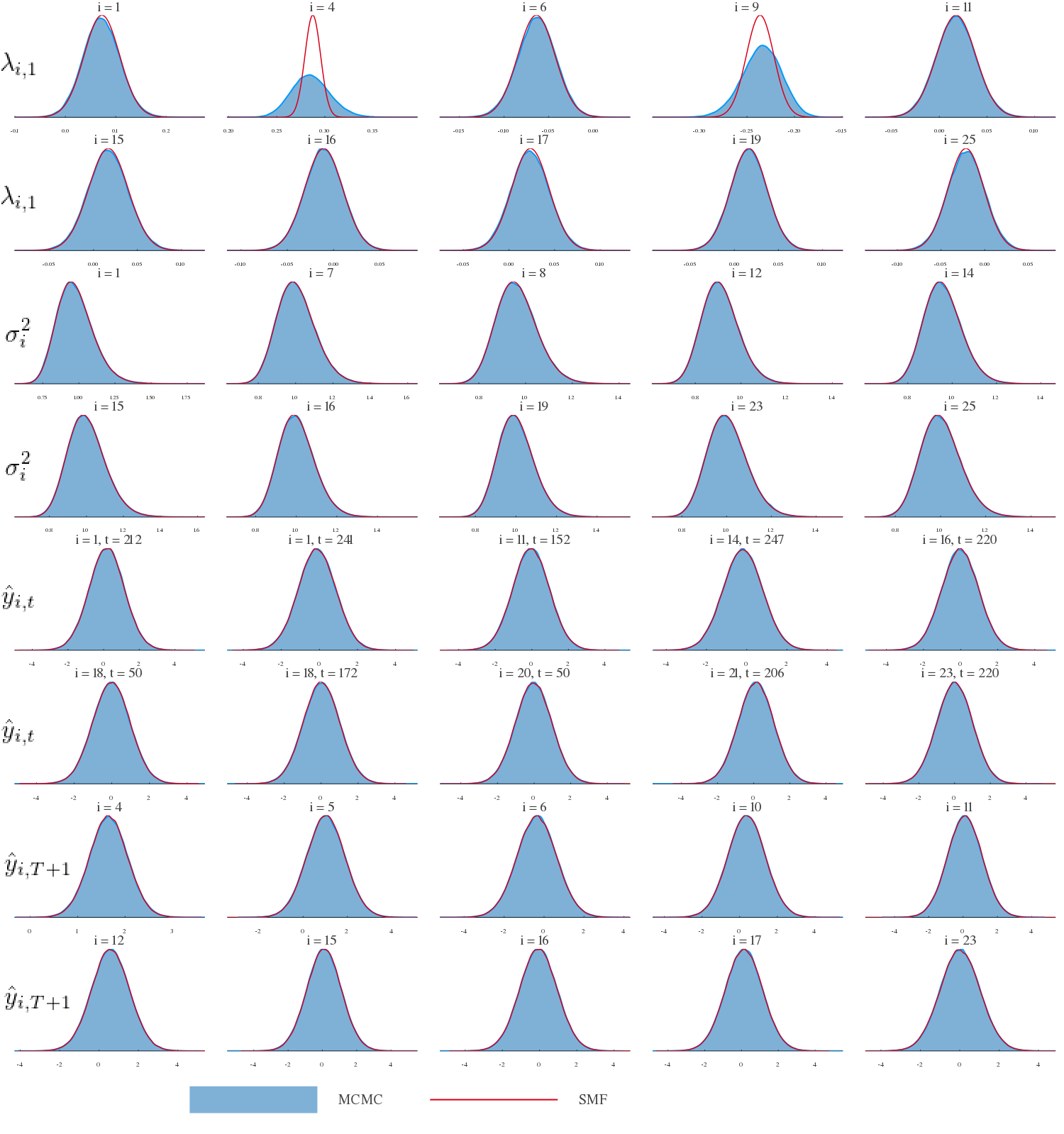}
    \label{SMFcomp}
\end{figure}

\pagebreak 
\begin{figure}[H]
    \centering
     \caption{Randomly selected parameters and predictions, MCMC-posterior and SMF-approximation, large DFM}
    \includegraphics[width=1\textwidth]{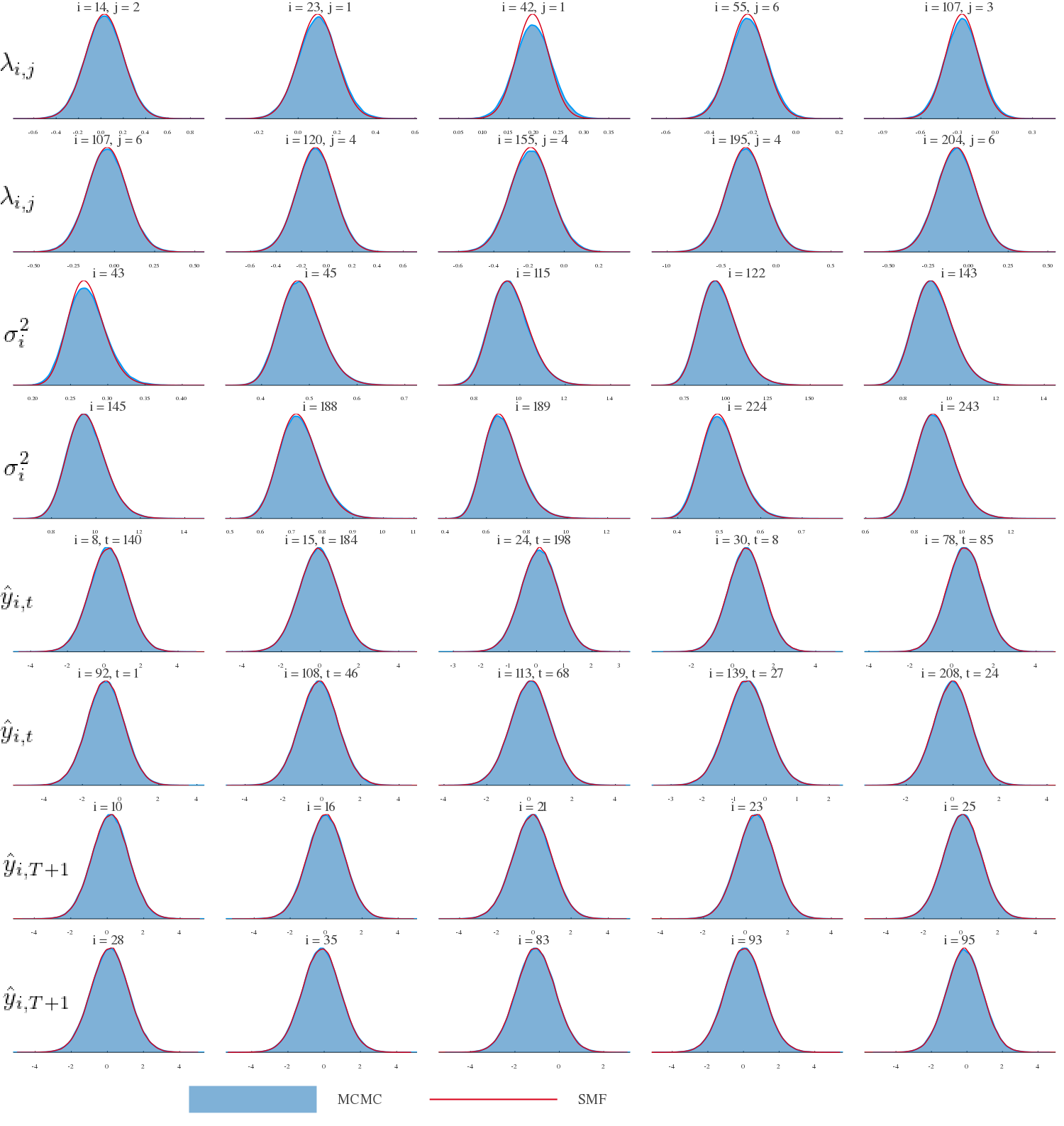}
    \label{fig:my_label1}
\end{figure}

\begin{figure}[H]
    \centering
    \caption[]{\tabular[t]{@{}l@{}} Evidence lower bound for algorithm iterations \\ \\\endtabular}
    \includegraphics[width=1\textwidth]{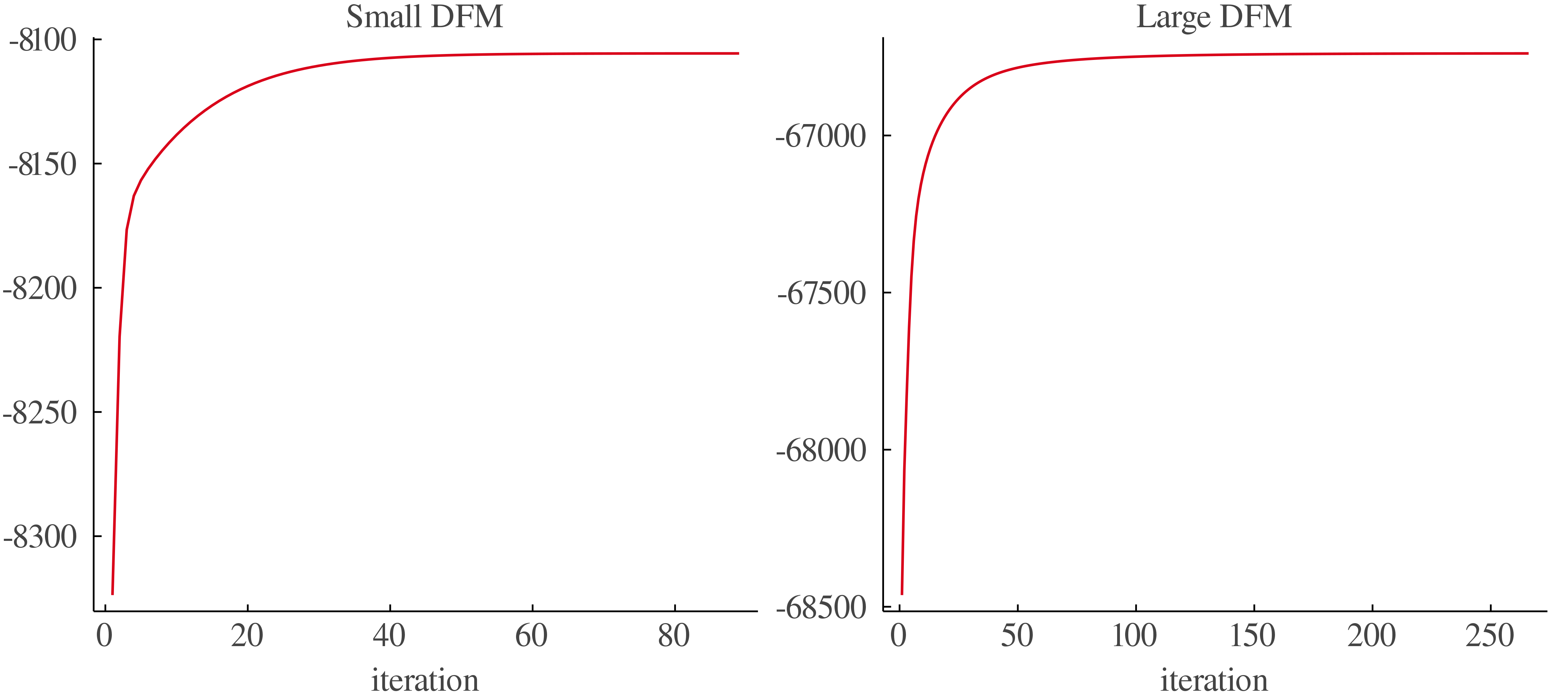}
    \label{ELBOfig}
\end{figure}
\end{appendices}
\newpage 
\setcounter{figure}{0}
\renewcommand\thefigure{\arabic{figure}}

\end{document}